\documentclass[useAMS,usenatbib]{mn2e}

\usepackage{aas_macros}  
\usepackage{graphicx}    
\usepackage{amsmath}     
\usepackage[             
  colorlinks=true,       
  urlcolor={blue},       
  linkcolor={black},
  citecolor={black},
  pdfborder={0 0 0}
]{hyperref}

\newcommand{\Prob}{\text{P}}           
\newcommand{\posterior}{\mathcal{P}}   
\newcommand{\lik}{\mathcal{L}}         
\newcommand{\prior}{\pi}               
\newcommand{\ev}{\mathcal{Z}}          
\newcommand{\data}{\mathcal{D}}        
\newcommand{\params}{\theta}           
\newcommand{\model}{\mathcal{M}}       

\newcommand{\programscript}[1]{{\sc #1}}              
\newcommand{\PolyChord}{\programscript{PolyChord}}     
\newcommand{\MultiNest}{\programscript{MultiNest}}     
\newcommand{\FORTRAN}{\programscript{FORTRAN}95}       
\newcommand{\openMPI}{\programscript{openMPI}}         
\newcommand{\CosmoMC}{\programscript{CosmoMC}}         
\newcommand{\CAMB}{\programscript{CAMB}}               
\newcommand{\CosmoChord}{\programscript{CosmoChord}}   
\newcommand{\ModeCode}{\programscript{ModeCode}}       
\newcommand{\ModeChord}{\programscript{ModeChord}}     

\newcommand{\smin}{\mathrm{min}}         
\newcommand{\smax}{\mathrm{max}}         
\newcommand{\slive}{\mathrm{live}}       
\newcommand{\sdead}{\mathrm{dead}}       
\newcommand{\sprocs}{\mathrm{procs}}     
\newcommand{\sdims}{\mathrm{dims}}       
\newcommand{\srepeats}{\mathrm{repeats}} 
\newcommand{\sshell}{\mathrm{shell}} 

\newcommand{\nlive}{n_\slive}       
\newcommand{\ndims}{n_\sdims}       
\newcommand{\nprocs}{n_\sprocs}     
\newcommand{\nlike}{N_\lik}         
\newcommand{\nrepeats}{n_\srepeats} 

\newcommand{\nhatx}{{\bf\hat{n}}}        
\newcommand{\nhat}[1]{\nhatx_{#1}} 
\newcommand{\bxx}{{\bf x}}                
\newcommand{\bmew}{{\bf \mu}}                
\newcommand{\bzero}{{\bf 0}}                
\newcommand{\byy}{{\bf y}}                
\newcommand{\bFF}{{\bf F}}                
\newcommand{\bx}[1]{\bxx_{#1}}         

\newcommand{\mean}[1]{{\left\langle{#1}\right\rangle}}
\newcommand{\bigO}[1]{{\sim\mathcal{O}{\left(#1\right)}}}

\newcommand{\citepos}[1]{\citeauthor{#1}'s (\citeyear{#1})}

\title[\PolyChord]{\PolyChord{}: next-generation nested sampling}

\author[W.J.~Handley et.\ al]
{W.J.~Handley$^{1,2}$\thanks{wh260@mrao.cam.ac.uk}, 
M.P.~Hobson$^1$\thanks{mph@mrao.cam.ac.uk} \& 
A.N.~Lasenby$^{1,2}$\thanks{a.n.lasenby@mrao.cam.ac.uk} 
\\
$^1$Astrophysics Group, 
Cavendish Laboratory, 
J.~J.~Thomson Avenue, 
Cambridge, 
CB3 0HE, 
UK 
\\
$^2$Kavli Institute for Cosmology,
Madingley Road,
Cambridge,
CB3 0HA,
UK
}

\date{Received 30 May 2015}
\pubyear{2015}
\pagerange{\pageref{firstpage}--\pageref{lastpage}} 

\begin{document}
\label{firstpage} 
\maketitle 

\begin{abstract}
  \PolyChord{} is a novel nested sampling algorithm tailored for high-dimensional parameter spaces. This paper coincides with the release of \PolyChord{} v1.3, and provides an extensive account of the algorithm. \PolyChord{} utilises slice sampling at each iteration to sample within the hard likelihood constraint of nested sampling. It can identify and evolve separate modes of a posterior semi-independently, and is parallelised using \openMPI{}. It is capable of exploiting a hierarchy of parameter speeds such as those present in \CosmoMC{} and \CAMB{}, and is now in use in the \CosmoChord{} and \ModeChord{} codes.  \PolyChord{} is available for download at: \url{http://ccpforge.cse.rl.ac.uk/gf/project/polychord/}
\end{abstract}


\section{Introduction}
\label{sec:introduction}
Over the past two decades, the quantity and quality of astrophysical and cosmological data has increased substantially.  In response to this, Bayesian methods have been increasingly adopted as the standard inference procedure.

Bayesian inference consists of {\em parameter estimation\/} and {\em model comparison}.  Parameter estimation is generally performed using Markov-Chain Monte-Carlo (MCMC) methods, such as the Metropolis--Hastings (MH) algorithm and its variants~\citep{Mackay}.  In order to perform model comparison, one must calculate the {\em evidence\/}: a high-dimensional integration of the likelihood over the prior density~\citep{Sivia}.  MH methods cannot compute this on a usable timescale, hindering the use of Bayesian model comparison in cosmology and astroparticle physics.

A contemporary methodology for computing evidences and posteriors simultaneously is provided by nested sampling~\citep{skilling2006}. This has been successfully implemented in the now widely adopted algorithm \MultiNest\,~\citep{MultiNest1,MultiNest2,MultiNest3}.  Modern cosmological likelihoods now involve a large number of parameters, with a hierarchy of speeds.  \MultiNest\ struggles with high-dimensional parameter spaces, and is unable to take advantage of this separation of speeds.  \PolyChord{} aims to address these issues, providing a means to sample high-dimensional spaces across a hierarchy of parameter speeds.

The layout of the paper is as follows:
Section~\ref{sec:bayesian_inference} is a general overview of parameter estimation and model selection in the context of Bayesian Inference.
In Section~\ref{sec:nested_sampling} we describe \citepos{skilling2006} nested sampling meta-algorithm.
We overview the historical implementations of nested sampling in Section~\ref{sec:iso_likelihood_sampling} and provide an account of \citepos{NealSlice} slice sampling technique.
We describe the \PolyChord{} algorithm in detail in Section~\ref{sec:polychord_algorithm} and demonstrate its efficacy on toy and cosmological problems in Section~\ref{sec:polychord_in_action}.  
Section~\ref{sec:conclusions} concludes the paper.
In addition we provide three appendices. Appendix~\ref{app:prior_tranformations} describes the procedure for implementing new prior distributions within the context of nested sampling.
Appendices~\ref{app:evidences}~\&~\ref{app:evidences_clusters} describe the mathematics of inferring evidences from the samples produced by nested sampling.

This paper is an extensive overview of our algorithm, which is now in use in several cosmological applications~\citep{planck2015-a24}. A briefer introduction can be found in~\cite{polychordletter}.

\PolyChord{} is available for download from the link at the end of the paper.

\section{Bayesian inference}
\label{sec:bayesian_inference}
In this section, we describe the key concepts of Bayesian inference necessary for understanding the utility of \PolyChord{}. 
For readers experienced in the field, this section serves to establish nomenclature and notation.
For a full discussion of Bayesian inference, we recommend~\cite{Sivia} or part IV of~\cite{Mackay}.

\subsection{Nomenclature}
\label{sec:nomenclature}
Scientific theory is concerned with the construction of predictive models in the context of some dataset $\data$.
A typical model $\model$ contains a set of variable parameters $\params_\model$. One may use $\model$ to calculate the probability of observing the data given a specific parameter choice:
\begin{equation}
  \Prob(\data|\params_\model,\model)\equiv\lik.
  \label{eqn:lik_def}
\end{equation}
This distribution on $\data$ is termed the {\em likelihood\/} $\lik$. From a Bayesian standpoint a model must also specify our initial degree of belief on the parameters $\params_\model$:
\begin{equation}
  \Prob(\params_\model|\model) \equiv \prior,
  \label{eqn:prior_def}
\end{equation}
This distribution on $\params_\model$ is termed the {\em prior\/} $\prior$. Typically this is a parametric distribution which quantifies our initial assumptions on the scale and spread of the parameters\footnote{Common examples include a uniform distribution between two bounds, or a Gaussian distribution with specified mean and variance.}.

The likelihood~(\ref{eqn:lik_def}) is conditioned on a set of chosen values for the model parameters $\params_\model$. One may marginalise out the dependence on $\params_\model$ by integrating over the prior distribution:
\begin{equation}
  \Prob(\data|\model) \equiv \ev = \int  \Prob(\data|\params_\model,\model)\Prob(\params_\model|\model)\:d\params_\model.
  \label{eqn:ev_int}
\end{equation}
This quantity is termed the {\em evidence\/} $\ev$, or {\em marginalised likelihood}, and gives the probability of observing the data $\data$, conditioned on the model $\model$. Suppressing explicit dependence on the model, the evidence computation can be written as:
\begin{equation}
  \ev = \int \lik(\params)\prior(\params)\:d\params.
  \label{eqn:evidence}
\end{equation}

\subsection{Parameter estimation}
\label{sec:param_est}

If the prior has been specified, Bayes theorem allows us to invert the conditioning in equation~(\ref{eqn:lik_def}) and find the {\em posterior\/} $\posterior$ by combining the likelihood, prior and evidence:
\begin{equation}
  \Prob(\params_\model|\data,\model) = \frac{\Prob(\data|\params_\model,\model) \Prob(\params_\model|\model)}{\Prob(\data|\model)},
  \label{eqn:bayes_theorem}
\end{equation}
which is schematically written as:
\begin{equation}
  \posterior = \frac{\lik \times \pi }{\ev }.
  \label{eqn:bayes_theorem_abbrv}
\end{equation}
This describes how our initial knowledge $\prior$ of the parameters updates to $\posterior$ in light of the data $\data$.
Calculation of the posterior $\posterior(\theta)$ is the domain of {\em parameter estimation}, and in high dimensions is best performed by sampling the space with a Markov-Chain Monte-Carlo approach (MCMC). Examples include Metropolis--Hastings, Gibbs sampling and Slice sampling. For the most part, the evidence $\ev$ is ignored during such calculations, and one works with an unnormalised posterior ${\posterior\propto\lik\times\prior}$.

\subsection{Model comparison}
\label{sec:model_comp}

Of equal importance in scientific investigation is {\em model comparison}. Typically one has multiple competing models ${\{\model_1,\model_2,\cdots\}}$, each with their own parameters and assumptions. The data $\data$ are able to decide on the relative merits of each of these models via Bayes theorem:
\begin{align}
  \Prob(\model_i|\data) &= \frac{\Prob(\data|\model_i)\Prob(\model_i)}{\Prob(\data)}, \\
  &= \frac{\ev_i\pi_i}{\sum_j \ev_j\pi_j}.
\end{align}
In contrast to parameter estimation, the evidences of each model $\ev_i$ take the leading role in model comparison. One typically will choose uniform priors on the models, ${\pi_i\equiv\Prob(\model_i)= \mathrm{const}}$, and then choose to use the model with the highest evidence. However, when evidences are similar in magnitude, the correct Bayesian approach is to make inferences by marginalising over all models considered. If there is a common derived parameter $y$, with marginalised posterior $\Prob(y|\data,\model_i)$ then one may produce the fully marginalised posterior:
\begin{equation}
  \Prob(y|\data) = \frac{\sum_i\Prob(y|\data,\model_i)\ev_i\pi_i}{\sum_j\ev_j\pi_j}.
\end{equation}
This fully Bayesian approach has been historically under utilised due to the difficulties in computing the evidence numerically from the integral~(\ref{eqn:evidence}).

\section{Nested sampling}
\label{sec:nested_sampling}
\begin{figure}
  \centering
  \includegraphics[width=\columnwidth]{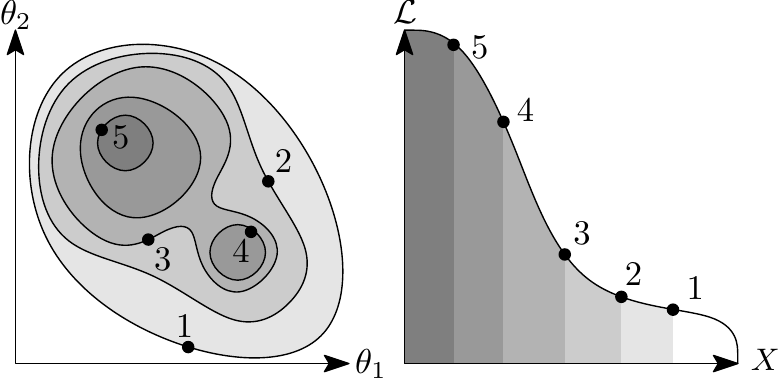}
  \caption{%
    The nested sampling volume transformation.
    Left: five iso-likelihood contours of a two-dimensional multi-modal likelihood function $\lik(\params)$. Each contour encloses some fraction of the prior $X$, indicated by colour.
    Right: Likelihood $\lik$ as a function of the volume $X$ enclosed by the contour. The evidence is the area under this curve.\label{fig:prior_volume}
  }
\end{figure}
\PolyChord{} falls into a category of sampling algorithms known as {\em nested sampling}. In order to explain the advances that \PolyChord{} has made, it is first necessary to describe the nested sampling meta-algorithm. Readers familiar with the theory may skip to Section~\ref{sec:iso_likelihood_sampling}.

Computing the evidence~(\ref{eqn:evidence}) typically involves an integral over a high-dimensional parameter space, only a small fraction of which contributes to $\ev$. The size and position of the region surrounding the peak(s) will not be known {\em a priori}, and in high dimensions is hard to find (see Figure~\ref{fig:prior_volume}).  

Algorithms need to be able quickly to compress the parameter space from the prior onto the posterior. In order to perform parameter estimation it needs to produce samples from the posterior, and to perform model comparison it should be able to calculate the evidence. 
Nested sampling~\citep{skilling2006} offers a means of doing all of these tasks simultaneously.

\subsection{Compressing the space}
\label{sec:comp_space}
Nested sampling maintains a population of $\nlive$ {\em live points\/} within a region of the parameter space. These points are sequentially updated so that the region that they occupy contracts around the peak(s) of the posterior. 

One begins by sampling $\nlive$ points from the prior distribution $\prior(\params)$. At iteration $i$, the point with the lowest likelihood $\lik_i$ is deleted, and then replaced by a new point. The new point is drawn from the prior, subject to the constraint that its likelihood is greater than $\lik_i$.

The fraction of the prior contained within an iso-likelihood contour $\lik(\params) = \lik$ is denoted the {\em prior volume\/}:
\begin{equation}
  X(\lik) = \int_{\lik(\params)>\lik} \prior(\params) d\params.
  \label{eqn:prior_volume}
\end{equation}
Since the live points are always drawn uniformly from $\prior(\params)$, at iteration $i$ the volume containing the live points will contract on average by a factor of ${\nlive/(\nlive+1)}$. Initially the prior volume is $1$, so at iteration $i$:
\begin{equation}
  \mean{X_i} = {\left( \frac{\nlive}{\nlive+1} \right)}^i \approx e^{-i/\nlive}.
  \label{eqn:X_full}
\end{equation}
The live points thus compress the prior {\em exponentially}. As the nested sampling run progresses, one is left with a sequence of discarded points (termed {\em dead points\/}). Each dead point will have a set of parameter values $\params_i$, a likelihood $\lik_i$ and an estimated prior volume $X_i$.

\subsection{Evidence estimation}
\label{sec:ev_es}
We can use the dead and live points to estimate the evidence. By differentiating the prior volume~(\ref{eqn:prior_volume}), we may re-write the evidence calculation~(\ref{eqn:evidence}) as an integral over a single variable:
\begin{equation}
  \ev = \int_0^1 \lik(X)dX.
  \label{eqn:evidence_short}
\end{equation}
This is detailed graphically in Figure~\ref{fig:prior_volume}. We may thus estimate the evidence by quadrature:
\begin{equation}
  \ev \approx \sum_{i\in\sdead} w_i \lik_i,
  \label{eqn:quadrature}
\end{equation}
where for simplicity we take $w_i = X_{i-1}-X_{i}$. Of course, this is only an estimate, since we are inferring the mean values $\mean{X_i}$ from the sampling procedure. One may however estimate the error in our inference, the full details of which can be found in Appendix~\ref{app:evidences}.

\subsection{Parameter estimation}
\label{sec:param_es}
Nested sampling can also perform parameter estimation by using the dead and live points as samples from the posterior, provided that the $i$th point is given the importance weighting:
\begin{equation}
  p_i = \frac{w_i\lik_i}{\ev},
  \label{eqn:importance_weighting}
\end{equation}
where $w_i$ is the prior volume of the shell in which point $i$ was sampled. 

\subsection{Algorithm termination}
\label{sec:alg_term}
\begin{figure}
  \centering
  \includegraphics[width=\columnwidth]{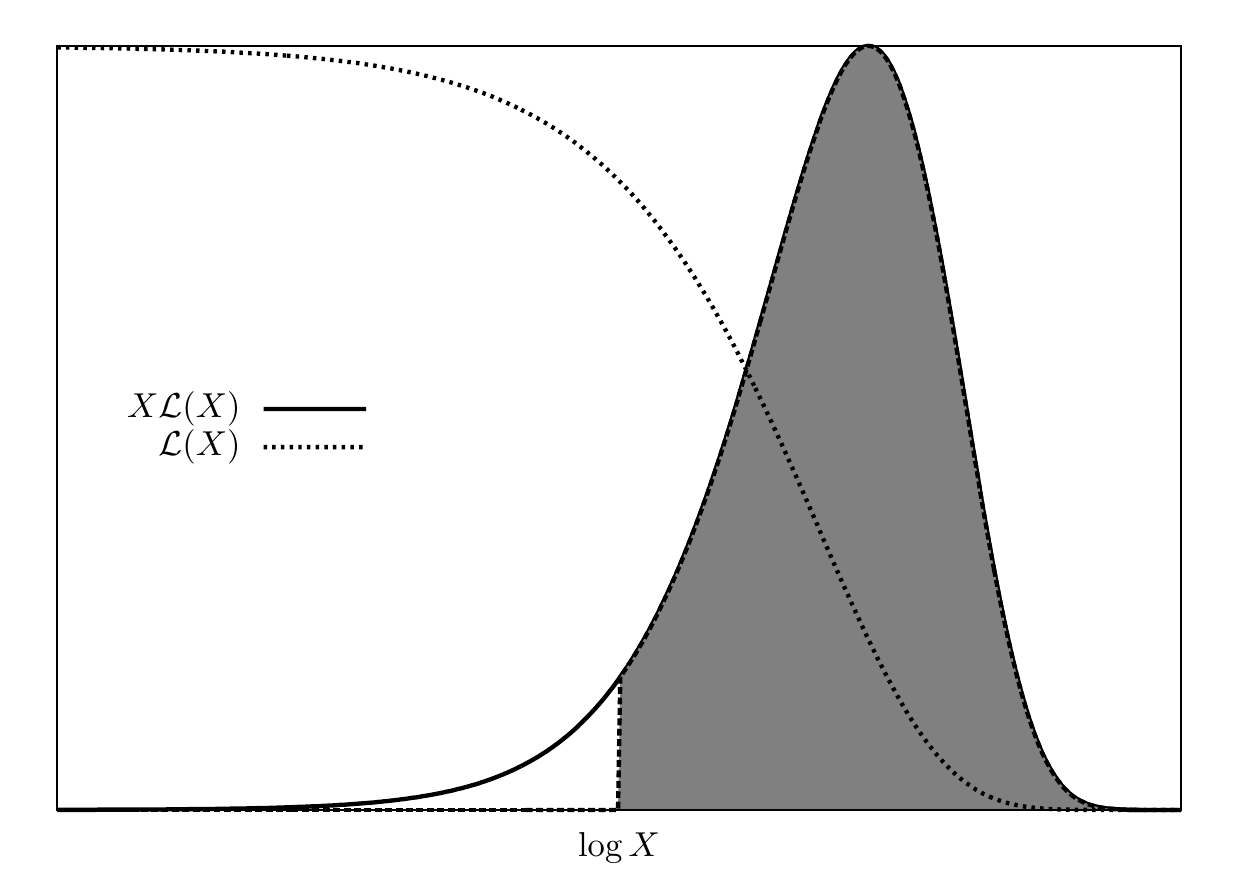}
  \caption{%
    Plot of a generic likelihood as a function of the prior volume $\lik(X)$. In high dimensions, the likelihood is only visible if plotted against $\log X$ (dashed curve). However, the evidence is better visualised by plotting $X\log(X)$ (solid curve). The area under the solid curve corresponds to the evidence. The magnitude of the solid curve is proportional to the importance weighting. Nested sampling proceeds from high to low volumes. After some time, the live points no longer contribute significantly to the evidence, and the algorithm terminates at this point.\label{fig:gaussian_weight}
  }
\end{figure}
As nested sampling proceeds, the likelihoods $\lik_i$ monotonically increase, but the weights $w_i$ monotonically decrease. This results in a peak in importance weights~(\ref{eqn:importance_weighting}) that can be seen in Figure~\ref{fig:gaussian_weight}. We terminate the algorithm once the remaining posterior mass (white region) left in the live points is some small fraction of the currently calculated evidence (dark region). The posterior mass left in the live points at iteration $i$ can be estimated by:
\begin{equation}
  \ev_\slive \approx \mean{\lik}_\slive X_i,
  \label{eqn:live_evidence}
\end{equation}
where the average is taken over the live points. Since this is typically an underestimate at early times, this will not cause premature termination.

\subsection{The unit hypercube}
\label{sec:unit_hypercube}
Each iteration of nested sampling requires one to sample from the prior (subject to a hard likelihood constraint). 
Typically, priors are defined in terms of simple analytic functions such as uniform or Gaussian distributions, and may be sampled using  inverse transform sampling. 

In the one-dimensional case, this amounts to converting a uniform random variable (which are easy to generate) into a variable sampled from a general distribution $f(\theta)$. One first finds its cumulative distribution function (CDF):
\begin{equation}
  F(\theta) = \int\limits_{-\infty}^\theta f(\theta^\prime) d\theta^\prime,
\end{equation}
computes the inverse of the CDF, and then applies this function to a uniform random variable  $x\sim U(0,1)$ to generate a variable $\theta = F^{-1}(x)$, which is distributed according to $f(\theta)$. 
In the general $D$-dimensional case, one calculates $D$ conditional distributions $\{F_i:i=1\ldots,D\}$ by marginalising over parameters $\theta_j, j>i$, and conditioning on $j<i$:
\begin{equation}
  F_i(\theta_i|\theta_{i-1},\ldots,\theta_1) = \int\limits_0^{\theta_i} f_i(\theta_i^\prime|\theta_{i-1},\ldots,\theta_1) d\theta_i^\prime,
\end{equation}
where:
\begin{equation}
  f_i(\theta_i|\theta_{i-1},\ldots,\theta_1) 
  =
  \frac{%
    \int f_i(\params) d\theta_{i+1}\ldots d\theta_{N}
  }{%
    \int f_i(\params) d\theta_{i}\ldots d\theta_{N}
  }.
\end{equation}
This generates a set of relations sequentially transforming $D$ uniform random variables $\{x_i\}$ into $\{\theta_i\}$ distributed according to $f(\params)$.

In many cases, the prior $\prior(\params)$ is separable, and the above equations are easily calculated. For sections of the parameters which are not separable, the calculation can become more involved. We include a few demonstrations of this procedure in Appendix~\ref{app:prior_tranformations}.

Nested sampling can thus be performed in the unit $D$-dimensional hypercube, $\bxx\in {[0,1]}^D$, defining a new likelihood function via $\lik(\params)= \lik(\bFF^{-1}(\bxx))$. This has numerous advantages, the first being that one only needs to be able to generate uniform random variables in $[0,1]$. The second is more subtle; it is more natural to define a distance metric in the unit hypercube than in the physical space. Unit hypercube variables all have the same dimensionality: probability.

\section{Sampling within an iso-likelihood contour}
\label{sec:iso_likelihood_sampling}
Now that the nested sampling meta-algorithm has been described, we briefly review the various instantiations that exist, and introduce \PolyChord{} as an algorithm utilising slice sampling at each iteration to generate new live points.

The most challenging aspect of nested sampling is drawing a new point from the prior subject to the hard likelihood constraint $\lik>\lik_i$. This may be done in a variety of ways, and distinguishes the various historical implementations.

\subsection{Previous Methods}
\label{sec:previous_methods}
For some problems, the iso-likelihood contour is known analytically, allowing one to construct a sampling procedure specific to that problem. This is demonstrated by~\cite{Keeton}, and can be useful for testing nested sampling's theoretical behaviour. In most cases, however, the likelihood contour is unknown a-priori, so a more numerical approach must be taken.

The \MultiNest\ algorithm~\citep{MultiNest1,MultiNest2,MultiNest3} samples by using the live points to construct a set of intersecting ellipsoids which together aim to enclose the likelihood contour, and then samples by rejection sampling within the ellipsoids. Whilst being an excellent algorithm for modest numbers of parameters, any rejection sampling algorithm has an exponential scaling with dimensionality that eventually emerges.

An alternative approach (the one initially envisaged by Skilling) is to sample with the hard likelihood constraint using a Markov-Chain based procedure. One makes several steps according to some proposal distribution until one is satisfied an independent sample is produced. This has significant advantages over a rejection-based approach, the most obvious being that the scaling with dimensionality is polynomial rather than exponential. In rejection sampling, points are drawn until one is found within the likelihood contour (often with extremely low efficiency). Using a Markov-chain approach however, (correlated) points are continually generated within the contour, until one is happy that a sample independent from the initial seed has been generated. These ``intra-chain points'' which we term {\em phantom points\/} have the potential to provide a great deal more information.

A traditional Metropolis--Hastings (MH) or Gibbs sampling approach may be utilised, but in general such algorithms are ill-suited to sampling from a hard likelihood constraint without a significant amount of tuning of a proposal matrix. This is examined in section 6 of~\cite{MultiNest1}.

Galilean (Hamiltonian) sampling~\citep{GalileanNestedSampling,Betancourt2011} improves upon the traditional MH sampler by using proposal points generated by reflecting off iso-likelihood contours. This however requires gradients to be calculated, and can become inefficient if the step size is chosen incorrectly, or if the contour has a shape which is difficult to `step back into'

Diffusive nested sampling~\citep{DiffusiveNestedSampling} is an alternative and promising variation on~\citepos{skilling2006} algorithm, which utilises MCMC to explore a mixture of nested probability distributions. Since it is MCMC based, it scales well with dimensionality. In addition, it can deal with multimodal and degenerate posteriors, unlike traditional MCMC\@. It does however have multiple tuning parameters.

\subsection{Slice sampling}
\label{sec:slice_sampling}
\begin{figure}
  \centerline{%
    \includegraphics{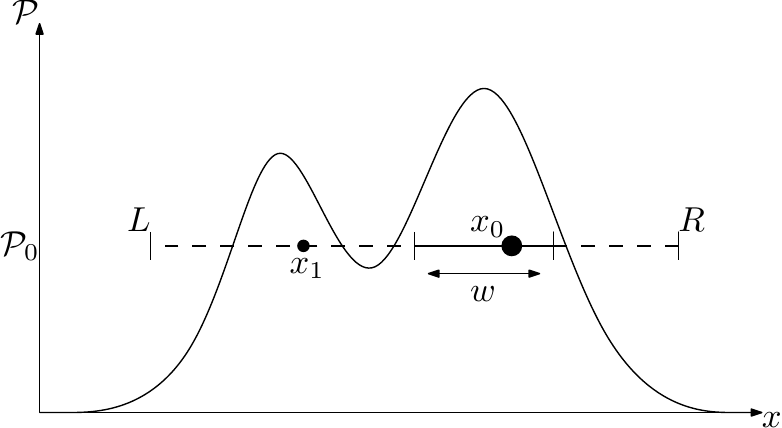}
  }

  \caption{Slice sampling in one dimension. 
    Given a probability level (or slice) $\posterior_0$, slice sampling samples within the horizontal region defined by $\posterior>\posterior_0$. 
    From an initial point $x_0$ within the slice ($\posterior(x_0)>\posterior_0$), a new point $x_1$ is generated within the slice with a distribution $P(x_1|x_0)$.
    External bounds are first set on the slice $\hat{L}<x_0<\hat{R}$ by uniformly expanding a random initial bound of width $w$ until they lie outside the slice (Neal terms this the {\em stepping out\/} procedure). 
    $x_1$ is then sampled uniformly within these bounds.  
    If $x_1$ is not in the slice, then $\hat{L}$ or $\hat{R}$ is replaced with $x_1$, ensuring that $x_0$ is still within the slice.
    This procedure is guaranteed to generate a new point $x_1$, and satisfies detailed balance $P(x_0|x_1) = P(x_1|x_0)$. Thus, if $x_0$ is drawn from a uniform distribution within the slice, so is $x_1$.\label{fig:1d_slice}
  }
\end{figure}

We have found that a Markov-Chain based procedure utilising ~\citepos{NealSlice} slice sampling at each step is well suited to sampling uniformly within an iso-likelihood contour.
Radford Neal initially proposed slice sampling as an effective methodology for generating samples numerically from a given posterior $\posterior(\params)$. One first chooses a `slice' (or probability level) $\posterior_0$ uniformly within $[0,\posterior_\smax]$. One then samples uniformly within the $\params$-region defined by $\posterior(\params)>\posterior_0$. The similarity with the iso-likelihood contour sampling required by nested sampling should be clear. In the one-dimensional case, he suggests the sampling procedure detailed in Figure~\ref{fig:1d_slice}.

This procedure for sampling within a likelihood bound is ideal for nested sampling. It samples uniformly with minimal information: an initial bound size $w$, and a point $x_0$ that is within the contour. In general $w$ must be chosen so that it is roughly the size of the bound, but if one overestimates it then the bounds will contract exponentially. Indeed, one may consider this as being equivalent to a prior space compression~(\ref{eqn:X_full}) with $\nlive=\ndims=1$. As a starting point, one may use one of the live points, which is already uniformly sampled. Since the procedure above satisfies detailed balance, this will produce a point which is also uniformly sampled within the iso-likelihood contour.

In higher dimensions,~\cite{NealSlice} suggests a variety of MCMC-like methods. The simplest of these is implemented by sampling each of the parameter directions in turn. Since each one-dimensional slice requires $\bigO{\text{a few}}$ likelihood calculations, the number of likelihood calculations required scales linearly with dimensionality. Multi-dimensional slice sampling has many of the benefits of a traditional MH approach, and uses a proposal distribution which is much more efficient at sampling a hard likelihood constraint.

\cite{SystemsBio} have already applied this procedure to nested sampling. This works exceptionally well for cases in which the parameters are non-degenerate. However, this becomes inefficient in the case of correlated parameters, or curving degeneracies.

\section{The \PolyChord{} algorithm}
\label{sec:polychord_algorithm}

\PolyChord{} implements several novel features compared to~\citepos{SystemsBio} slice-based nested sampling.  
It utilises slice sampling in a manner that uses the information present in the live and phantom points to deal with correlated posteriors. 
\PolyChord{} also uses a general clustering algorithm that identifies and evolves separate modes of the posterior semi-independently, and infers local evidence values.  
In addition, it has the option of implementing fast-slow parameters, which is extremely effective in its combination with \CosmoMC{}~\citep{cosmomc}. 
This is termed \CosmoChord, which may be downloaded from the link at the end of the paper.

 The algorithm is written in \FORTRAN{} and parallelised using \openMPI{}.  It is optimised for the case where the dominant cost is the generation of a new live point.  This is frequently the case in astrophysical applications, either due to high dimensionality, or to costly likelihood evaluation.  

\begin{figure*}
  \centerline{%
    \includegraphics[width=\textwidth]{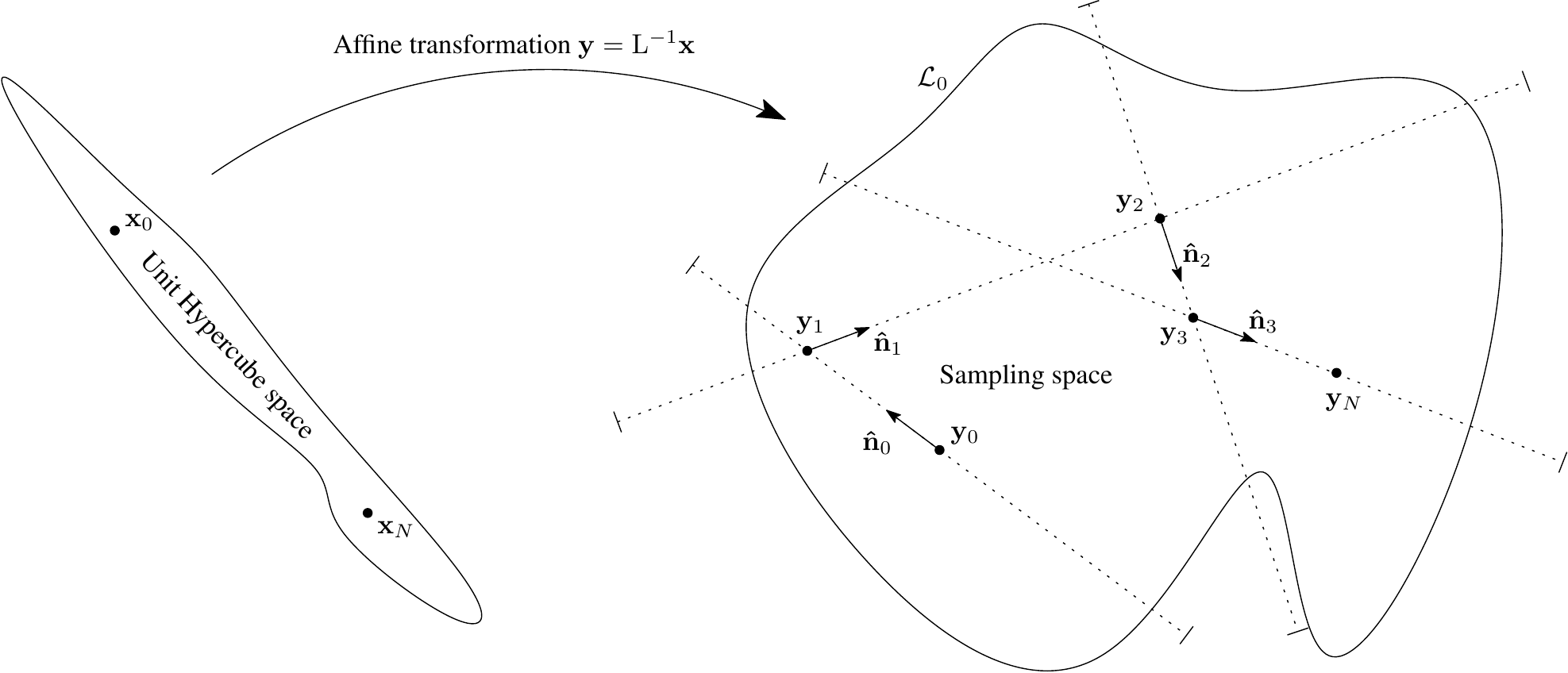}
}
\caption{%
  Slice sampling in $D$ dimensions. 
  We begin by ``whitening'' the unit hypercube by making a linear transformation which turns a degenerate contour into one with dimensions $\bigO{1}$ in all directions. 
  This is a linear skew transformation defined by the inverse of the Cholesky decomposition of the live points' covariance matrix. 
  We term this whitened space the {\em sampling space}. 
  Starting from a randomly chosen live point $\bx{0}$, we pick a random direction and perform one-dimensional slice sampling in that direction (Figure~\protect\ref{fig:1d_slice}), using $w=1$ in the sampling space. 
  This generates a new point $\bx{1}$ in $\bigO{\text{a few}}$ likelihood evaluations. 
  This process is repeated $\bigO{\ndims}$ times to generate a new uniformly sampled point $\bx{N}$ which is decorrelated from $\bx{0}$.\label{fig:Nd_slice}
}
\end{figure*}

\subsection{Multi-dimensional slice sampling}
\label{sec:multi_slice}
At each iteration $i$ of nested sampling, we generate a new randomly sampled point within the iso-likelihood contour $\lik_i$ by our variant of $D$-dimensional slice sampling.
Slice sampling is performed in the unit hypercube with hypercube coordinates denoted in bold ($\bxx$).

At each iteration $i$ of the nested sampling algorithm, one of the live points is chosen at random as a start point for a new chain with hypercube coordinate $\bx{0}$. We then make a one-dimensional slice sampling step (Figure~\ref{fig:1d_slice}) with initial width $w$ in a random direction $\nhat{0}$ chosen from a probability distribution $\Prob(\nhatx)$. This generates a new point $\bx{1}$ which is uniformly sampled in the unit hypercube, but is correlated to $\bx{0}$. This process is repeated $\nrepeats$ times, with $\bx{j-1}$ forming the start point for a slice along $\nhat{j-1}$ to produce $\bx{j}$. This procedure is illustrated in the right hand half of Figure~\ref{fig:Nd_slice}.

Since the probability of drawing $\bx{j}$ from $\bx{j-1}$ is the same as the probability of drawing $\bx{j-1}$ from $\bx{j}$, this procedure satisfies detailed balance. Thus, the resulting chain will ergodically be uniformly distributed within the iso-likelihood contour. This also applies to multi-modal posteriors, with the chance of jumping out a mode being equal to the chance of jumping back in.

The length of the chain $\nrepeats$ should be large enough so that the final point of the chain is decorrelated from the start point. 
This final point may now be considered to be a new uniformly sampled point from the prior distribution subject to the hard likelihood constraint. The intermediate points are saved and stored as phantom points. Whilst phantom points are correlated, they are useful in providing additional information and posterior points.

There are several elements of this which are left undetermined, namely the probability distribution $\Prob(\nhatx)$, the initial width $w$, and the chain length $\nrepeats$. These issues are addressed in the next section.

\subsection{Contour whitening}
\label{sec:cont_white}
In order to determine an optimal $\Prob(\nhatx)$ and $w$, an algorithm will need some knowledge of the contour in which the chain is progressing. This information can be supplied by the set of live and phantom points which are already uniformly distributed within the contour. We use the sample covariance matrix of the live and phantom points as a proxy for the size and shape of the contour.

Uniformly sampled points remain uniformly sampled under an affine transformation. The covariance matrix is used to construct an affine transformation which ``whitens'' the contour. Sampling is then performed in this whitened space, which we term the {\em sampling space}.
In the sampling space, the contour has size $\bigO{1}$ in every direction. This means that one may choose the initial step size as $w=1$.

To transform from $\bxx$ in the unit hypercube to $\byy$ in the sampling space we use the relation:
\begin{equation}
  \mathrm{L}^{-1}\bxx =  \byy,
  \label{eqn:cholesky}
\end{equation}
where $\mathrm{L}$ is the Cholesky decomposition of the covariance matrix $\Sigma = \mathrm{L} \mathrm{L}^{T}$.
This is illustrated further in Figure~\ref{fig:Nd_slice}.

Working in the sampling space our choice of $\Prob(\nhatx)$ is inspired by the default choice of \CosmoMC{}~\citep{LewisFastSlow}. Here, a randomly oriented orthonormal basis is chosen, and these directions are chosen in a random order. Once a basis is exhausted, a new basis is chosen. This approach satisfies detailed balance, and mixes rapidly.

The choice of $\nrepeats$ is slightly harder to justify. We find that for distributions with roughly convex contours $\nrepeats\bigO{\ndims}$ is sufficient, with the constant of proportionality being $2$---$6$. For more complicated contour shapes, one may require much larger values of $\nrepeats$. 

This procedure has the advantage of being dynamically adaptive, and requires no tuning parameters. However, this ``whitening'' process is ineffective for pronounced curving degeneracies. This will be discussed in detail in Section~\ref{sec:gaussian_shells}.

\subsection{Clustering}
\label{sec:clustering}
Multi-modal posteriors are a challenging problem for any sampling algorithm. ``Perfect'' nested sampling (i.e.\ the entire prior volume enclosed by the iso-likelihood contour is sampled uniformly) in theory solves multi-modal problems as easily as uni-modal ones. In practice however, there are two issues.

First, one is limited by the resolution of the live points. If a given mode is not populated by enough live points, it runs the risk of ``dying out''. Indeed, a mode may be entirely missed if the density of live points is too low. In many cases, this problem can be alleviated by increasing the number of live points.

Second, and more importantly for \PolyChord{}, the sampling procedure may not be appropriate for multi-modal problems. We ``whiten'' the unit hypercube using the covariance matrix of live points. For far-separated modes, the covariance matrix will not approximate the dimensions of the contours, but instead falsely indicate a high degree of correlation.  It is therefore essential for our purposes to have \PolyChord{} recognise and treat modes appropriately.

This methodology splits into two distinct parts:
  (i) recognising that clusters are there, and
  (ii) evolving the clusters semi-independently.

\subsubsection{Cluster recognition}
\label{sec:clustering_recognition}
Any cluster recognition algorithm can be substituted at this point.  One must take care that this is not run too often, or one runs the risk of adding a large overhead to the calculation.  In practice, checking for clustering every $\bigO{\nlive}$ iterations is sufficient, since the prior will have only compressed by a factor $e$.  We encourage users of \PolyChord{} to experiment with their own preferred cluster recognition, in addition to that provided and described below. 

It should be noted that the live points of nested sampling are amenable to most cluster recognition algorithms for two reasons.  First, all clusters should have the same density of live points in the unit hypercube.  Second, there is no noise (i.e.\ outside of the likelihood contour there will be no live points). Many clustering algorithms struggle when either of these two conditions is not satisfied.

We therefore choose a relatively simple variant of the $k$-nearest neighbours algorithm to perform cluster recognition.  If two points are within one another's $k$-nearest neighbours, then these two points belong to the same cluster.  We iterate $k$ from $2$ upwards until the clustering becomes stable (the cluster decomposition does not change from one $k$ to the next).  If sub-clusters are identified, then this process is repeated on the new sub-clusters.

\subsubsection{Cluster evolution}
\label{sec:clustering_evolution}
An important novel feature comes from what one does once clusters are identified. 

First, when spawning from an existing live point, the whitening procedure is now defined by the covariance matrix of the live points within that cluster. This solves the issue detailed above.

Second, by choosing a random initial live point as a seed, \PolyChord{} would naively spawn live points into a mode with a probability proportional to the number of live points in that mode. In fact, what it should be doing is to spawn in proportion to the volume fraction of that mode. These should be the same, but the difference between these two ratios will exhibit random-walk like behaviour, and can lead to biases in evidence calculations, or worse, cluster death. Instead, one can keep track of an estimate of the volume in each cluster, and choose the mode to spawn into in proportion to that estimate. This methodology is documented in Appendix~\ref{app:evidences_clusters}. 

In addition to keeping track of local volumes, we may keep track of local evidences. At the moment of splitting, the existing evidence in the initial cluster is partitioned between the new sub-clusters. Upon algorithm completion, one is left with an estimate of the proportion of the evidence contained within each cluster, and thus a measure of the importance of the various modes. By partitioning the local evidences at cluster recognition, the local evidences will sum to give the total evidences, to within the error on our inference.

Thus, the point to be killed off is still the global lowest-likelihood point, but we control the spawning of the new live point into clusters by using our estimates of the volumes of each cluster. We call this `semi-independent', because it retains global information, whilst still treating the clusters as separate entities.

When spawning within a cluster, we determine the cluster assignment of the new point by which cluster it is nearest to. It does not matter if clusters are identified too soon; the evidence calculation will remain consistent.

\subsection{Parallelisation}
\label{sec:parallelisation}
\PolyChord{} is parallelised by \openMPI{} using a master-slave structure.  One master process takes the job of organising all of the live points, whilst the remaining ${\nprocs-1}$ ``slave'' processes take the job of finding new live points. This layout is optimised for the case where the dominant cost is the generation of a new live point due to the calculation of relatively expensive likelihoods.

When a new live point is required, the master process sends a random live point and the Cholesky decomposition to a waiting slave.  The slave then, after some work, signals to the master that it is ready and returns a new live point and the intra-chain points to the master.

\begin{figure}
  \centering
  \includegraphics[width=\columnwidth]{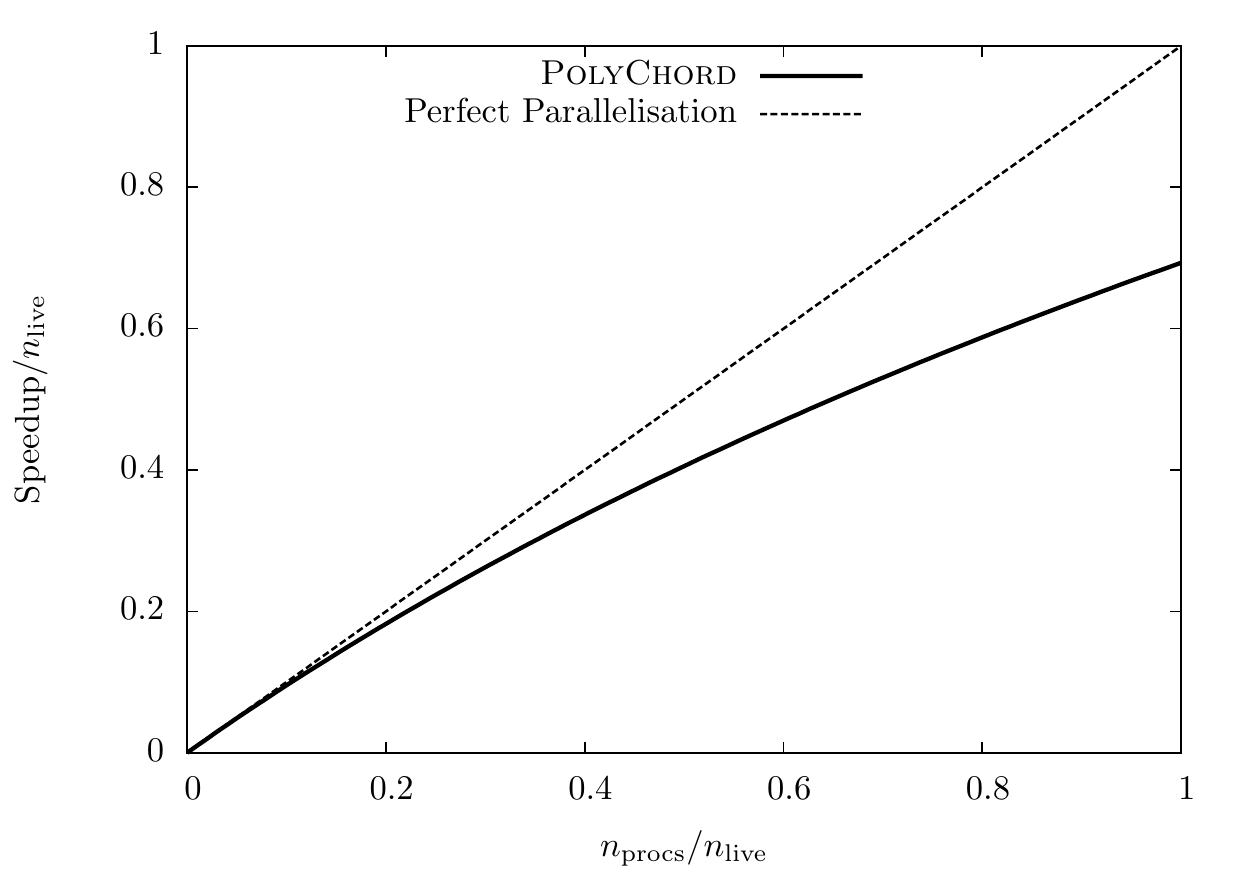}
  \caption{%
Parallelisation of \PolyChord{}. 
The algorithm parallelises nearly linearly, providing that $\nprocs<\nlive$. For most astronomical applications this is more than sufficient.\label{fig:parallel}}
\end{figure}

A point generated from an iso-likelihood contour $\lik_i$ is usable as a new live point for an iso-likelihood contour $\lik_j>\lik_i$, providing it is within both contours.  One may keep slaves continuously active, and discard any points returned which are not usable.  The probability of discarding a point is proportional to the volume ratio of the two contours, so if too many slaves are used, then most will be discarded.  The parallelisation goes as:
\begin{equation}
  \text{Speedup}(\nprocs) = \nlive\log\left[ 1 + \frac{\nprocs}{\nlive} \right],
  \label{eqn:parallel}
\end{equation}
and is illustrated in Figure~\ref{fig:parallel}. As a rule, \PolyChord{} parallelises almost linearly up to the number of live points, but from then on exhibits a law of diminishing returns. Since the number of live points is typically high $\bigO{500}$, this is more than sufficient for currently available \openMPI{} architectures, and certainly superior to the parallelisation of the standard Metropolis--Hastings algorithm.
\subsection{Posterior bulking}
\label{sec:posterior_bulking}
In addition to lending information on the scale and shape of a contour, phantom points can also be used as posterior samples. Correlations between samples are unimportant for the purposes of parameter estimation, providing one has enough to be well mixed. We may thus use the importance weighting detailed in~(\ref{eqn:importance_weighting}) with $w_i$ being set to the volume of the live-point shell which they occupy.

For high-dimensional cosmological applications, this results in a very large number ($\gg$GB) of posterior samples being produced, so \PolyChord{} thins these samples. From a user's perspective, one supplies a parameter which determines the fraction of phantom points to keep.

\subsection{Fast-slow parameters and \CosmoChord}
\label{sec:fast_slow}

In cosmological applications, likelihoods can exhibit a hierarchy of parameters in terms of calculation speed~\citep{LewisFastSlow}. Consequently, a likelihood may be quickly recalculated if one changes only a certain subset of the parameters. For \PolyChord{} it is very easy to exploit such a hierarchy. Our transformation to the sampling space is laid out so that if parameters are ordered from slow to fast, then this hierarchy is automatically exploited: a Cholesky decomposition, being a upper-triangular skew transformation, mixes each parameter only with faster parameters.

From a user's perspective, \PolyChord{} does this re-ordering in the hypercube automatically when provided with details of the hierarchy.

Further to this, one may use the fast directions to extend the chain length by many orders of magnitude. This helps to ensure an even mixing of live points. \PolyChord{} automatically times likelihood calculation speeds, so the user just has to provide what fraction of time \PolyChord{} should be spending on each subset of the parameters, and the algorithm will oversample accordingly.

\subsection{Tuning parameters}
\label{sec:tuning_params}

From a user's perspective, the \PolyChord{} algorithm has two tuning paramaters: $\nlive$ and $\nrepeats$, which are detailed below.

The authors believe that these tuning parameters are fairly straightforward to set in comparison to existing algorithms. More importantly, the number of tuning parameters does not scale with the dimensionality of the problem. This is in contrast to Metropolis--Hastings and Gibbs sampling, which require a proposal matrix to be supplied\footnote{Proposal matrices may be learnt during run-time. However, this learning step can take a prohibitively long time and reduces the efficacy of these approaches.}.

There are also several other options controlling run time behaviour, such as the production of equally weighted posterior samples, whether or not to perform clustering and the production and use of files allowing \PolyChord{} to resume from a previous run. These are documented in the input files supplied with the code.

\subsubsection*{Resolution $\nlive$ }
This is a generic nested sampling parameter. $\nlive$ indicates the number of live points maintained throughout the algorithm. Increasing $\nlive$ causes nested sampling to contract more slowly in volume (equation~\ref{eqn:X_full}), and consequently sample the space more thoroughly. Thus, it can be thought of as a resolution parameter. Run time scales $\bigO{\nlive}$

If set too low, posterior modes may be missed. Increasing $\nlive$ increases the accuracy of the inference of $\ev$, since the evidence error scales $\bigO{\nlive^{-1/2}}$. 

\subsubsection*{Reliability $\nrepeats$}
This is a \PolyChord{} specific parameter. It corresponds to the length of the slice sampling chain used to generate a new live point. Increasing this parameter decreases the correlation between live points, and hence increases the reliability of the evidence inference. Posterior estimations, however, remain accurate even in the event of low $\nrepeats$.

Setting this too low can result in correlation between live points, and unreliable evidence estimates. Typically, setting this $\bigO{3\times\ndims}$ is sufficient, but for curving degeneracies one may need significantly longer chains. Run time scales $\bigO{\nrepeats}$. 

\section{\PolyChord{} in action}
\label{sec:polychord_in_action}
We aim to showcase \PolyChord{} as both a high-dimensional evidence calculator, and multi-modal posterior sampler. We begin by comparing its dimensionality scaling with \MultiNest{}. We then demonstrate its clustering capabilities in high dimensions, and on difficult clustering problems. \PolyChord{} is shown to perform well on moderately pronounced curving degeneracies, and its implementation in \CosmoMC{} is discussed.

\subsection{High-dimensional evidences}
\label{sec:hi_ev}

As an example of the strength of \PolyChord{} as a high-dimensional evidence estimator, we compare it to \MultiNest{} on a Gaussian likelihood in $D$ dimensions.  In both cases, convergence is defined as when the posterior mass contained in the live points is $10^{-2}$ of the total calculated evidence.  We set $\nlive=25D$, so that the evidence error remains constant with $D$. \MultiNest{} was run in its default mode with importance nested sampling and expansion factor $e=0.1$.  Whilst constant efficiency mode has the potential to reduce the number of \MultiNest{} evaluations, the low efficiencies required in order to generate accurate evidences negate this effect.

\begin{figure}
  \centering
  \includegraphics[width=\columnwidth]{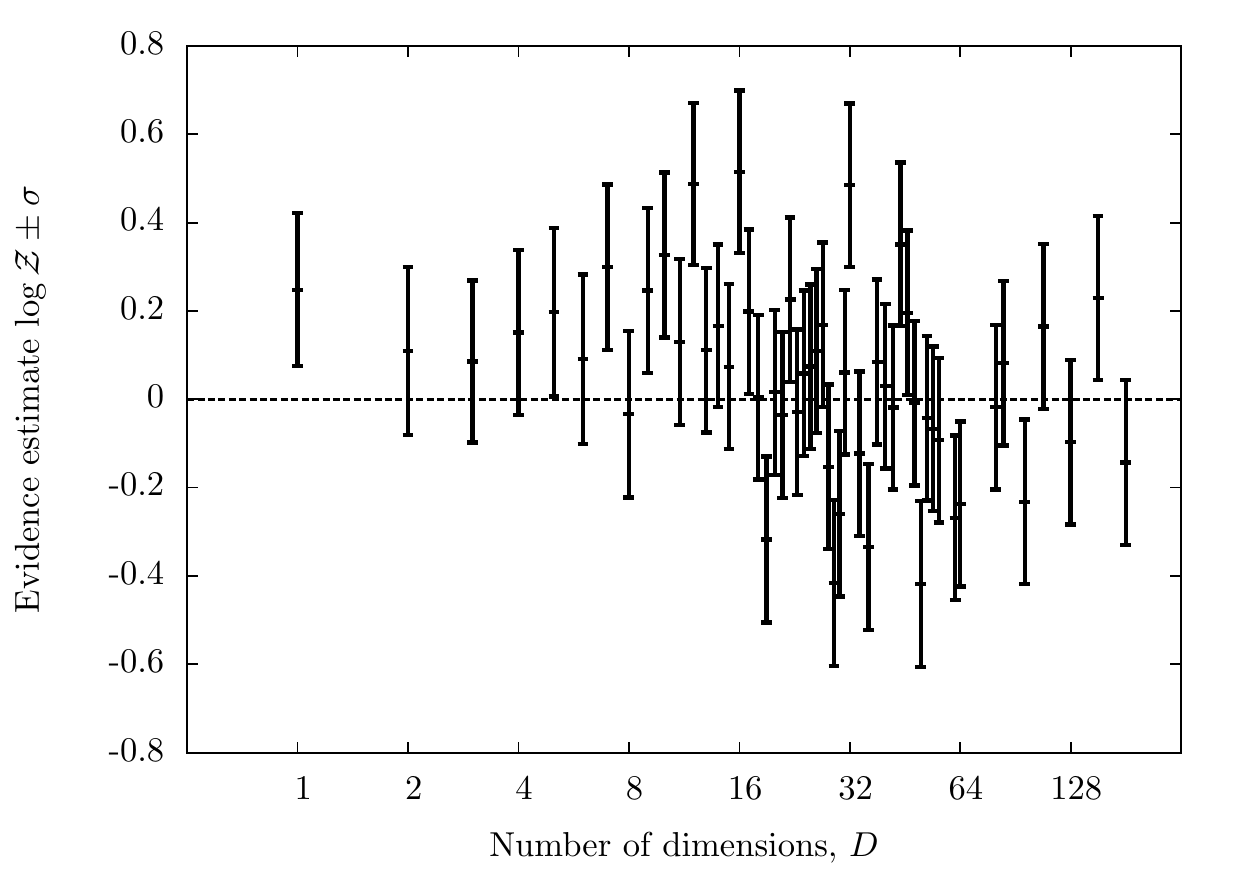}
  \caption{%
    Evidence estimates and errors produced by \PolyChord{} for a Gaussian likelihood as a function of dimensionality. The dashed line indicates the correct analytic evidence value.\label{fig:gaussian_evidences}
}
\end{figure}

With these settings, \PolyChord{} produces consistent evidence and error estimates with an error $\sim0.4$ log units (Figure~\ref{fig:gaussian_evidences}). Using importance nested sampling, \MultiNest{} produces estimates that are within this accuracy.

\begin{figure}
  \centering
  \includegraphics[width=\columnwidth]{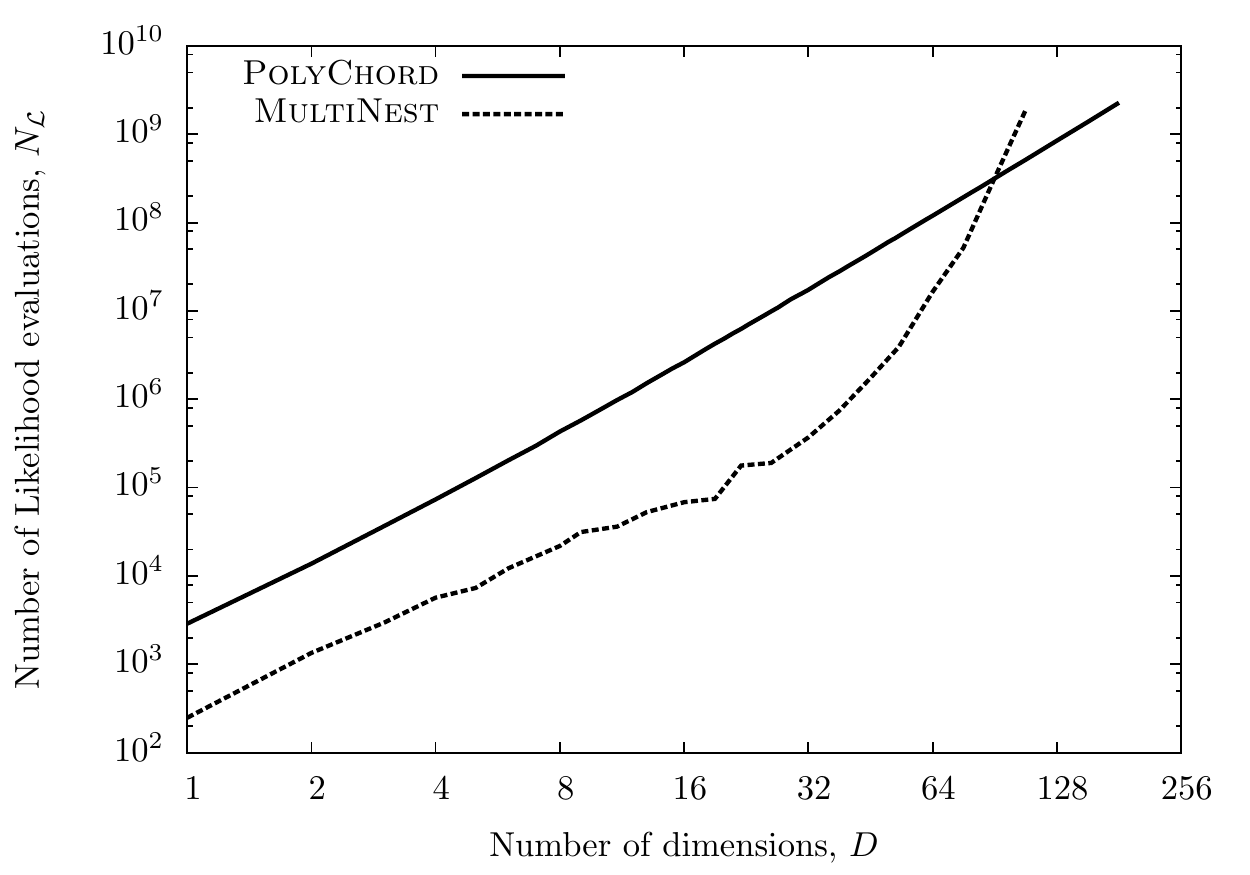}
  \caption{Comparing \PolyChord{} with \MultiNest{} using a
  Gaussian likelihood for different dimensionalities. \PolyChord{} has at worst $\nlike\bigO{D^3}$, whereas \MultiNest{} has an exponential scaling that emerges at high dimensions.\label{fig:gaussian}
}
\end{figure}

Figure~\ref{fig:gaussian} shows the number of likelihood evaluations $\nlike$ required to achieve convergence as a function of dimensionality $D$. 
Even on a simple likelihood such as this, \PolyChord{} shows a significant improvement over \MultiNest{} in scaling with dimensionality.  \PolyChord{} at worst scales as ${\nlike\bigO{D^3}}$, whereas \MultiNest{} has an exponential scaling which emerges in higher dimensions.
However, we must point out that a good rejection algorithm like \MultiNest{} will always win in low dimensions.

\subsection{Clustering and local evidences}
\label{sec:loc_ev}
To demonstrate \PolyChord{}'s clustering capability we report its performance on a ``Twin Peaks'' and Rastrigin likelihood.

\subsubsection{Twin peaks}
\label{sec:twin_peaks}
\PolyChord{} is capable of clustering posteriors in very high dimensions. We define a twin peaks likelihood as an equal mixture of two spherical Gaussians, separated by a distance of 10$\sigma$.

\PolyChord{} correctly identifies these clusters in arbitrary dimensions (tested up to $D=100$), providing that $\nlive$ and $\nrepeats$ are scaled in proportion to $D$. It calculates a global evidence that agrees with the analytic results. In addition, the local evidences correctly divide the peaks in proportion to their evidence contribution.

The results for a twin peaks likelihood are of an identical character to Figures~\ref{fig:gaussian_evidences}~\&~\ref{fig:gaussian}, and hence not included.

\subsubsection{Rastrigin function}
\label{sec:rastrigin}

\begin{figure}
  \centering
  \includegraphics[width=\columnwidth]{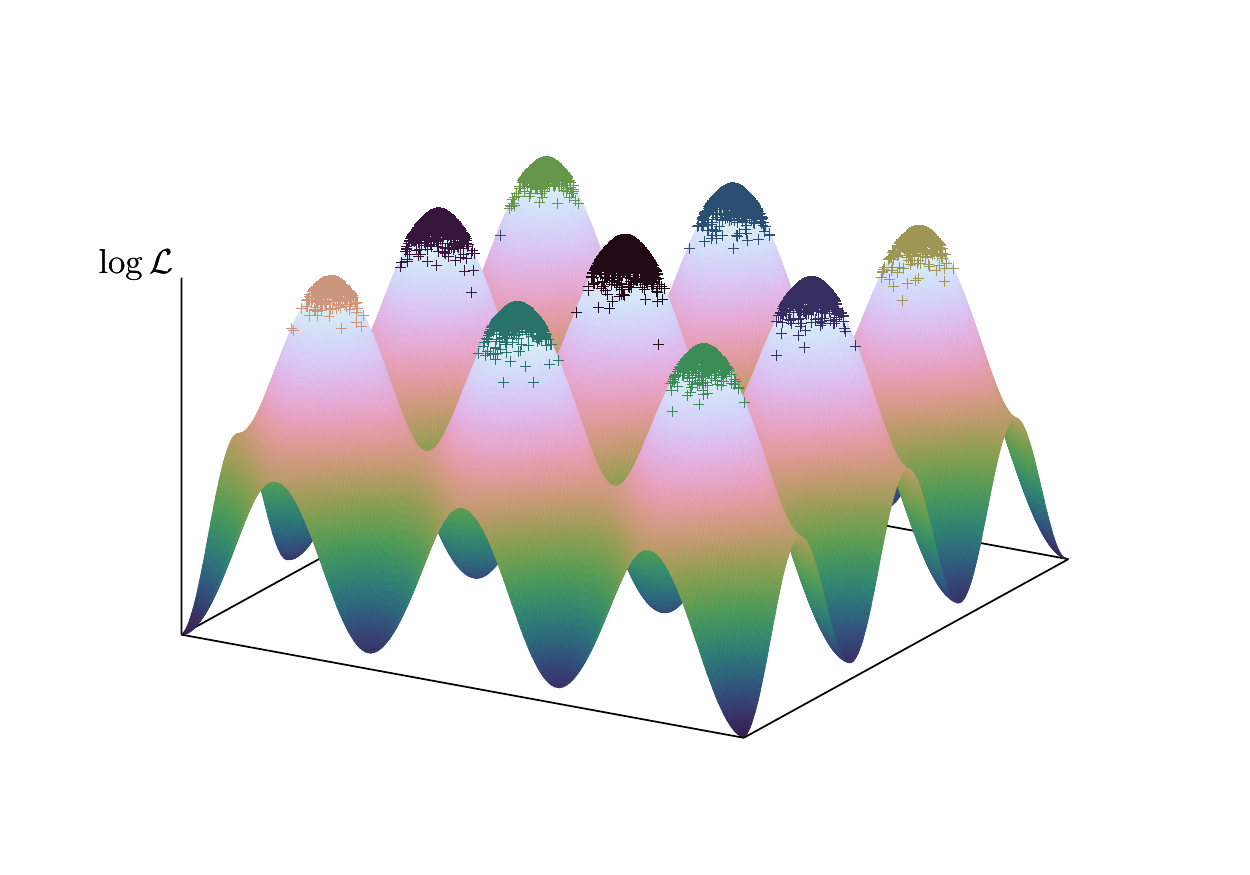}
  \caption{The two-dimensional Rastrigin $\log$-likelihood in the range ${[-1.5,1.5]}^2$. Within this region there are $8$ local maxima, and one global maximum at $(0,0)$. The clustered samples produced by \PolyChord{} are plotted on the $\log$-likelihood surface, with colours that indicating the separate clusters identified.\label{fig:rastrigin}}
\end{figure}

\begin{figure}
  \centering
  \includegraphics[width=\columnwidth]{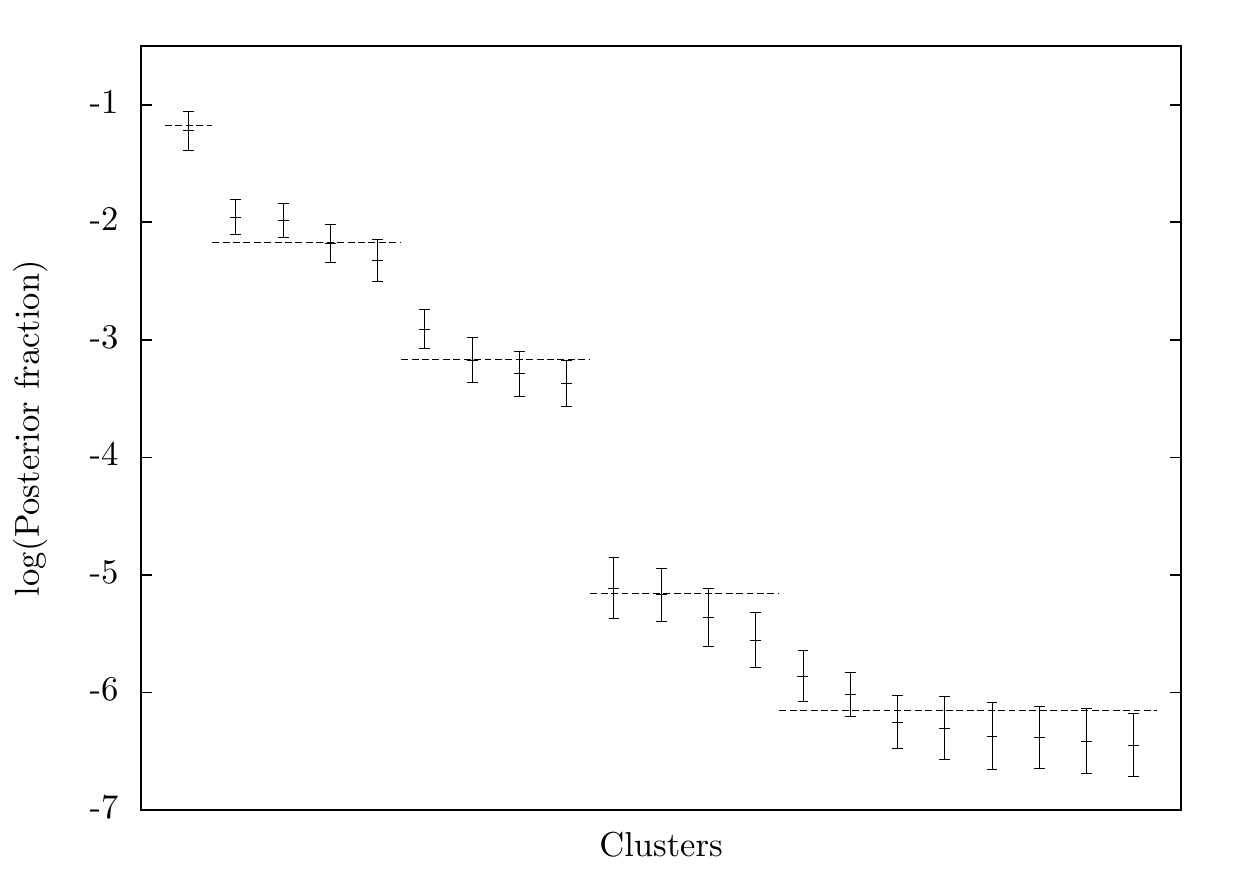}
  \caption{\PolyChord{} cluster identification for the Rastrigin function. \PolyChord{} identifies posterior modes and computes their local evidences, expressed here as a logarithmic fraction of  the total evidence in the mode. Dashed lines indicate the analytic results computed by a saddle point approximation at each of the peaks. As can be seen, \PolyChord{} reliably identifies the inner $21$ modes with increasing accuracy.\label{fig:rastrigin_data}}
\end{figure}

\PolyChord{}'s clustering capacity is very effective on complicated clustering problems as well. The $n$-dimensional Rastrigin test function is defined by:
\begin{align}
  f(\theta) &= A n + \sum\limits_{i=1}^n \left[\theta_i^2 - A\cos(2 \pi \theta_i) \right],
  \label{eqn:rastrigin_function}
  \\
  A&=10, \qquad \theta_i \in [-5.12,5.12]. \nonumber
\end{align}
This is the industry standard ``bunch of grapes'', the two-dimensional version of which is illustrated in Figure~\ref{fig:rastrigin}.
For our purposes, we will treat~(\ref{eqn:rastrigin_function}) as the negative log-likelihood so that $\lik(\theta) \propto \exp[-f(\theta)]$.
This is a stereotypically hard problem to solve, as many algorithms get stuck in local maxima.

We ran \PolyChord{} on a two-dimensional Rastrigin log-likelihood  with $\nlive=1000$ and $\nrepeats=6$. With these settings, \PolyChord{} calculates accurate evidence and posterior samples (Figure~\ref{fig:rastrigin}), and in addition correctly isolates and computes local evidences for the inner $21$ modes. Additional outer modes are also found, but these are combinations of lower modes due to their very low posterior fraction. Increasing the resolution parameter $\nlive$ further increases the number of modes identified.  Examples of clustered posterior samples are indicated in Figure~\ref{fig:rastrigin_data}, coloured using \citepos{cubehelix} `cubehelix'.

\subsection{Rosenbrock function}
\label{sec:rosenbrock}

\begin{figure}
  \centering
  \includegraphics[width=\columnwidth]{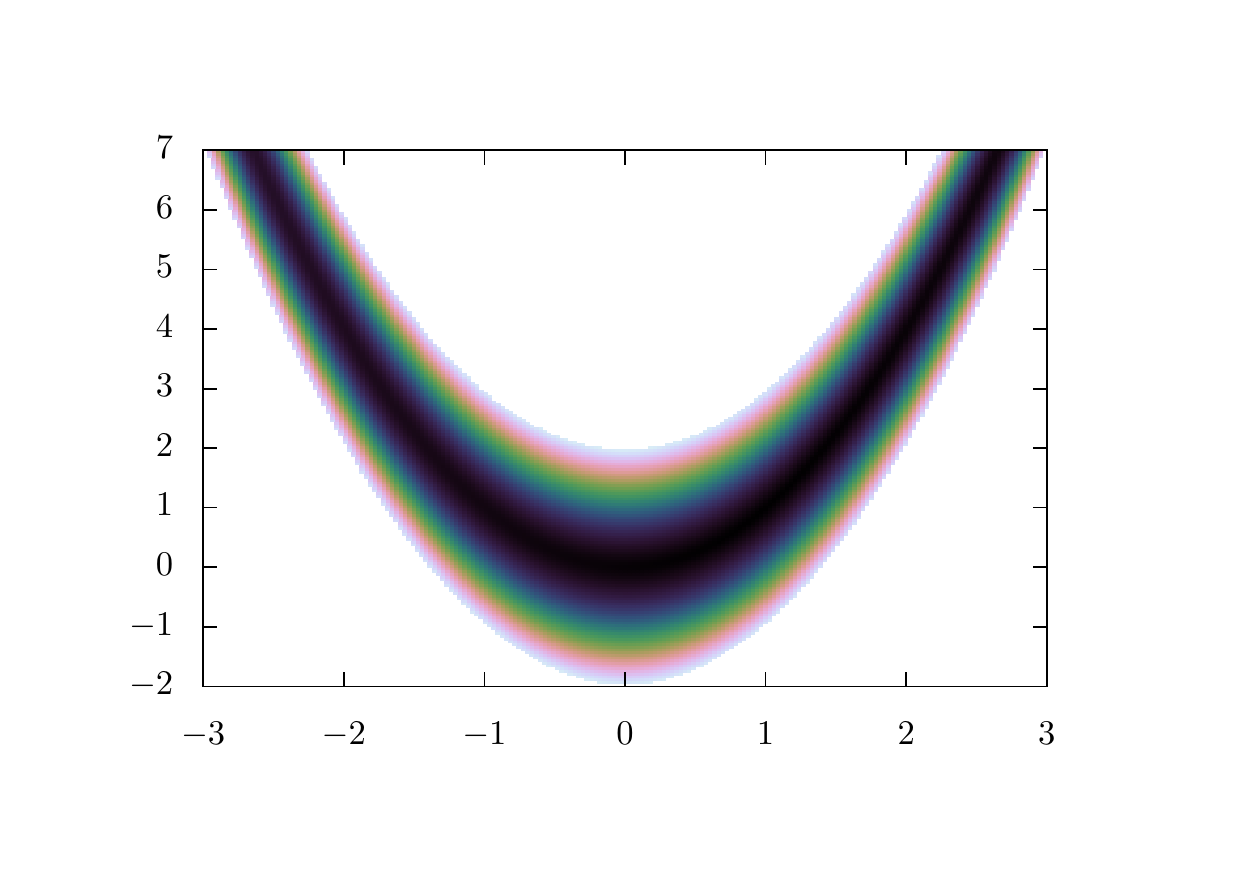}
  \caption{Density plot of the two-dimensional Rosenbrock function. The function exhibits a long, thin curving degeneracy, with a global maximum at $(1,1)$. \label{fig:rosenbrock_2d}}
\end{figure}

\begin{figure}
  \centering
  \includegraphics[width=\columnwidth]{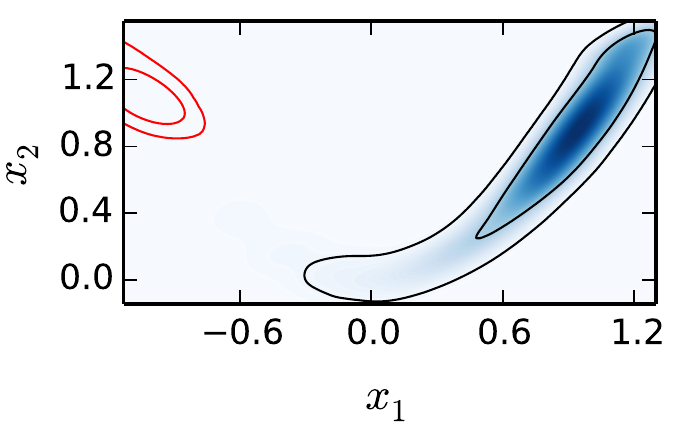}
  \caption{The four-dimensional Rosenbrock posterior, with $x_3$ and $x_4$ marginalised out. \PolyChord{} correctly identifies both the local (red) and global (blue) maxima.\label{fig:rosenbrock}}
\end{figure}

\PolyChord{} is also capable of navigating moderate curving degeneracies. 

The $n$-dimensional Rosenbrock function is defined by:
\begin{align}
  f(x) &= \sum\limits_{i=1}^{n-1}   {(a-x_i)}^2+ b {(x_{i+1} -x_i^2 )}^2,
    \label{eqn:rosenbrock}
    \\
    a&=1,\quad b=100,\quad x_i\in[-5,5],
\end{align}
the two-dimensional version of which is plotted in Figure~\ref{fig:rosenbrock_2d}. This is the industry standard ``banana'', as it exhibits an extremely long and flat curving degeneracy. We consider ${n=4}$, in which there is a global maximum at $(1,1,1,1)$ and a local maximum at $(-1,1,1,1)$, \PolyChord{} finds both of these (Figure~\ref{fig:rosenbrock}) and produces correct evidence estimations.

In higher dimensions, \PolyChord{} reliably finds the local and global maxima. The lack of an analytic evidence value for the Rosenbrock function prevents a verification of the evidence calculation.

\subsection{Gaussian shells}
\label{sec:gaussian_shells}

\begin{figure}
  \centering
  \includegraphics[width=\columnwidth]{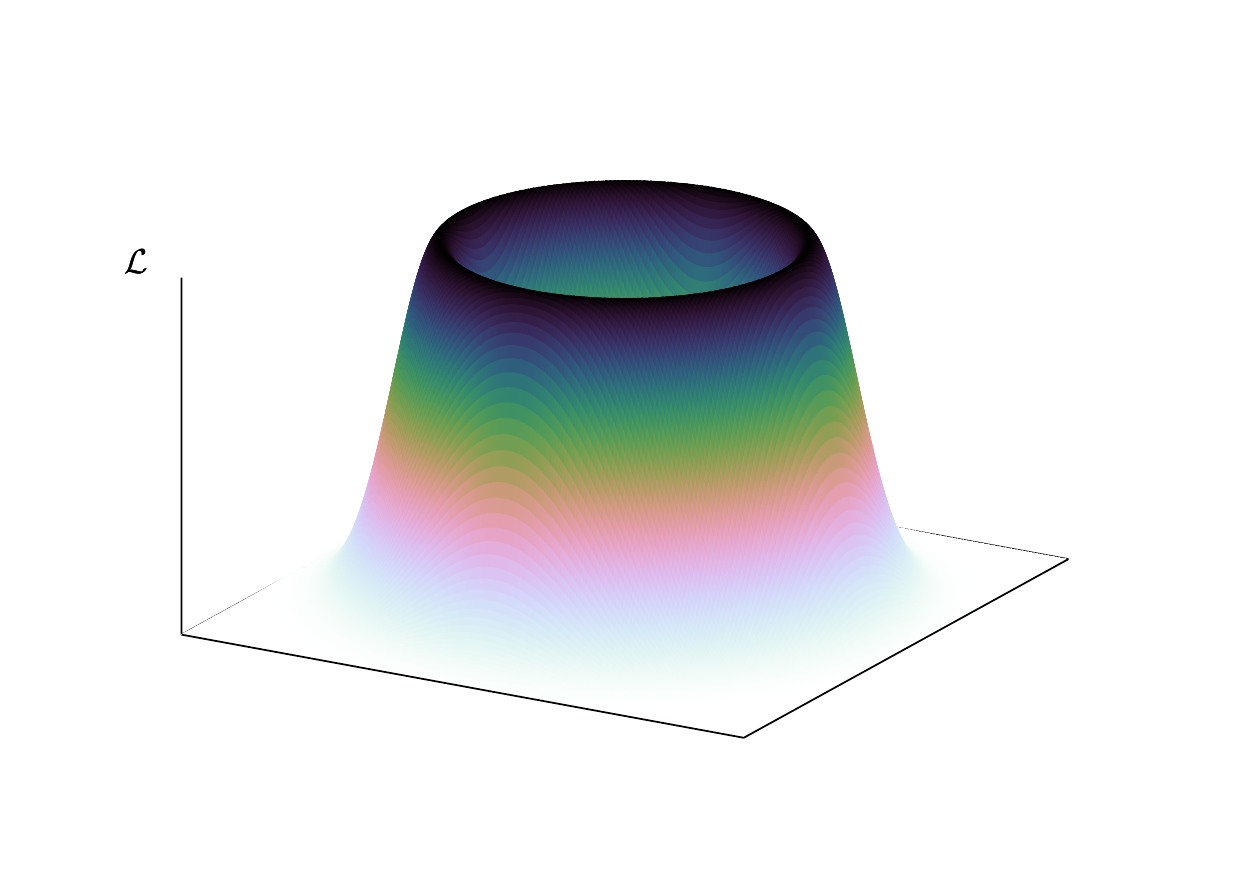}
  \caption{The two-dimensional Gaussian shell likelihood.\label{fig:gaussian_shell}}
\end{figure}

A ``Gaussian shell'' with mean $\bmew$, radius $r$ and width $w$ is defined as:
\begin{equation}
  \log\lik_\sshell(\bxx|\bmew,r,w) = A - \frac{{\left(\left|\bxx - \bmew\right|- r\right)}^2}{2w^2},
  \label{eqn:gaussian_shell}
\end{equation}
where $A$ is a normalisation constant that may be calculated using a saddle point approximation.
This likelihood is centered on some mean vector $\bmew$, and has a radial Gaussian profile with width $w$ at distance $r$ from this centre. This radial profile is then revolved around $\bmew$ to create a spherical shell-like likelihood. A two-dimensional version of this likelihood is indicated in Figure~\ref{fig:gaussian_shell}.

This distribution may be representative of likelihoods that one may encounter in beyond-the-Standard-Model paradigms in particle physics. In such models, the majority of the posterior mass lies in thin sheets or hypersurfaces through the parameter space.

\begin{figure}                                               
  \centering
  \includegraphics[width=\columnwidth]{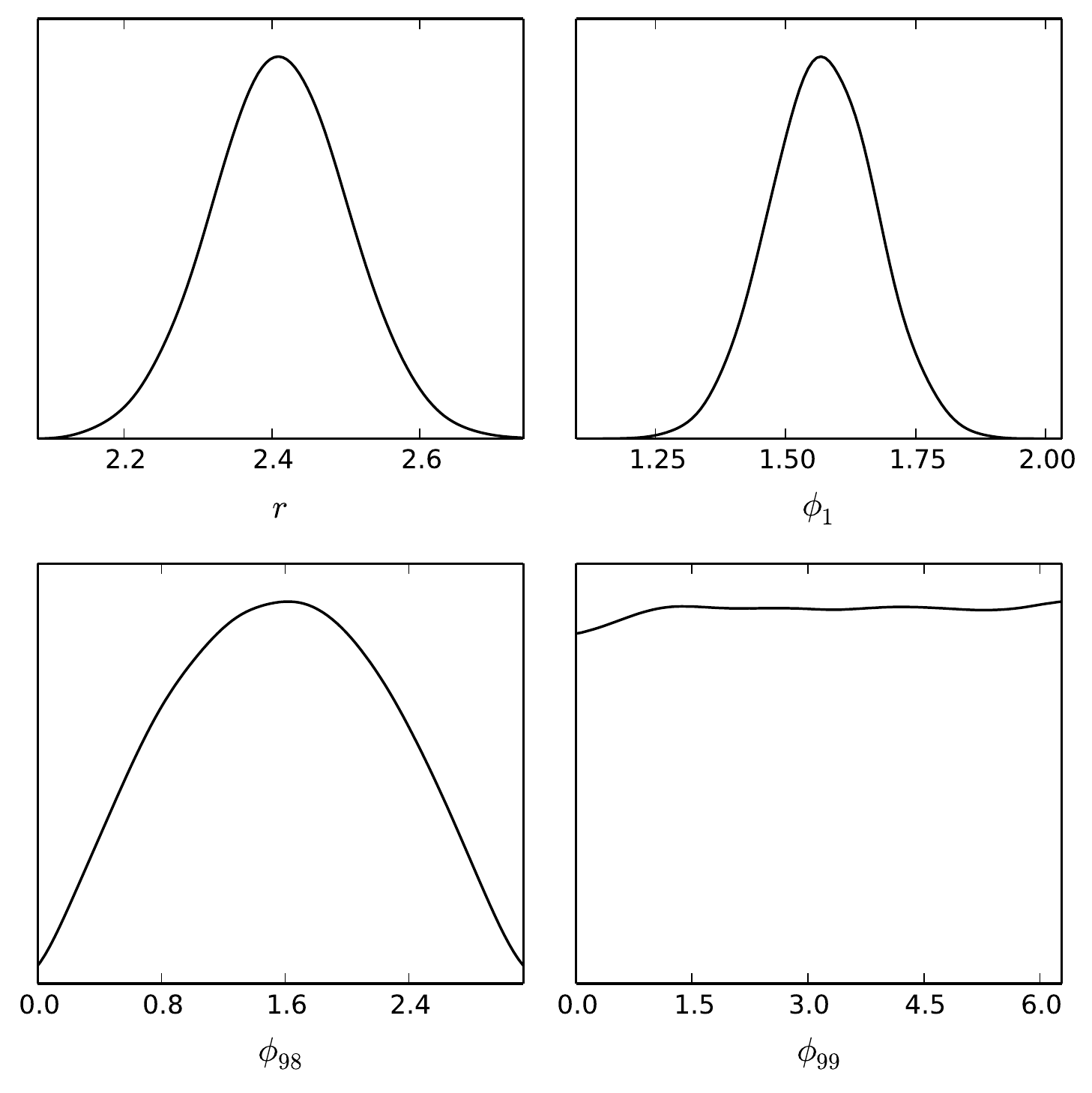}
  \caption{Posteriors produced by \PolyChord{} for a $n=100$-dimensional Gaussian shell, with width $w=0.1$, radius $r=2$, and center $\bmew=\bzero$. 
  Plotting the marginalised posteriors for the Cartesian sampling parameters ${\{x_1,\cdots,x_n\}}$ yields Gaussian distributions centered on the origin. To see the effectiveness of the sampler it is better to plot the sampling parameters in terms of $n$-dimensional spherical polar coordinates ${\{r,\phi_1,\cdots,\phi_{n-1}\}}$. Note that the polar coordinates are {\em derived parameters\/}, and that the sampling space still has the strong Gaussian shell degeneracy.
  In this case we can see that the radial coordinate has a Gaussian profile centered on $r_0 = r\times\frac{1}{2} {\left(1 + \sqrt{1 +  4 (n-1) {\left({w}/{r}\right)}^2}\right) }$ with width $w_0 = w{(1+(n-1){(w/r_0)}^2)}^{-1/2}$. 
  The azimuthal coordinate $\phi_{n-1}$ has a uniform posterior, and the other angular coordinates $\{\phi_i\}$ have posteriors defined by $\Prob(\phi_i ) \propto {\left(\sin\phi_i\right)^{n-i-1}}$.
\label{fig:gaussian_shell_posterior}
}
\end{figure}

Running \PolyChord{} on a $100$-dimensional Gaussian shell with $\nlive=1000$, $\nrepeats=200$ yields consistent evidences and posteriors, shown in Figure~\ref{fig:gaussian_shell_posterior}. 
                                                                  
Given that this problem is quoted as being ``optimally difficult'' \citep{MultiNest2}, the ease with which \PolyChord{} tackles this problem in high dimensions is worth explanation. In the two-dimensional case, it is clear that the posterior mass is concentrated in a very thin, curving region of the parameter space. However, as the dimensionality is increased, more and more of the $n$-sphere's volume is concentrated at the edge, and the thin characteristic of the degeneracy is lost. 

This may mean that the Gaussian shell is not a good proxy for a high-dimensional curving degeneracy. However, it could equally suggest that curving degeneracies become easier to navigate in higher dimensions. We can certainly conclude that a particle physics model with a proliferation of phases would be easier to navigate than one with a smaller number of phases.

\subsubsection{Twin Gaussian shells}
We finish our toy problems by combining the difficulties of multimodality (Section~\ref{sec:loc_ev}) and degeneracy, by mixing two twin Gaussian shells together:
\begin{equation}
  \lik(\bxx) \propto \lik_\sshell(\bxx|\bmew_1,r,w) + \lik_\sshell(\bxx|\bmew_2,r,w).
  \label{eqn:gaussian_shells}
\end{equation}
We choose $r=2$, $w=0.1$, and $\mu_1$ and $\mu_2$ are separated by $7$ units. With $\nlive=10\ndims$ and $\nrepeats=2\ndims$, \PolyChord{} successfully computes the local and global posteriors and evidences up to $D=100$, and reliably identifies the two modes. The comparison of run times with \MultiNest{} recovers a similar pattern to Figure~\ref{fig:gaussian}, although in our experience, the \MultiNest{} parameters require some tuning to ensure that evidences are calculated correctly when $\ndims>30$.

\subsection{\CosmoChord}
\label{sec:cosmochord}

\begin{figure*}
  \centering
  \includegraphics[width=\textwidth]{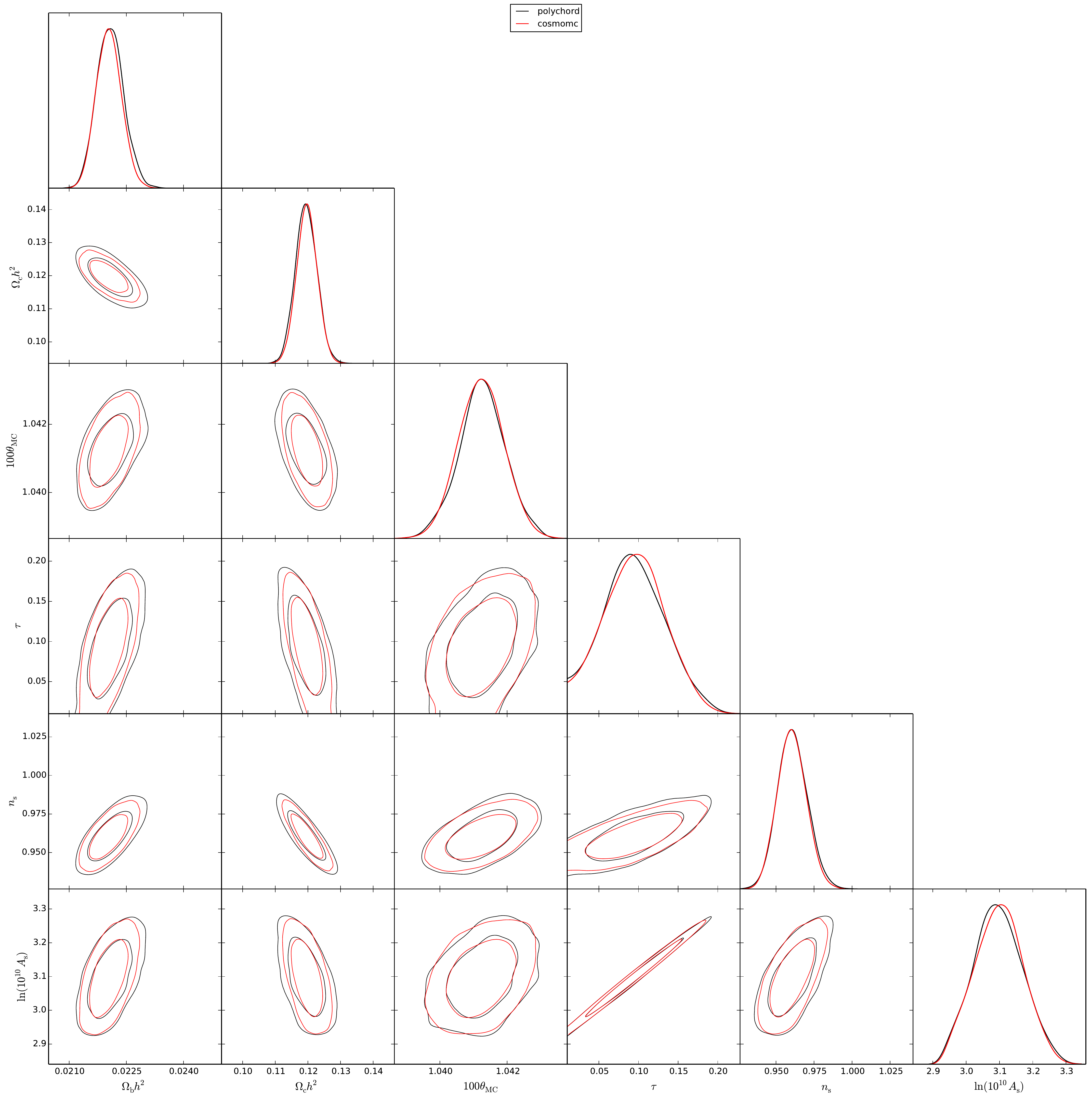}
  \caption{\CosmoChord{} (red) vs.\ \CosmoMC{} (black). We use the 2013 {\tt CAMSPEC}+{\tt commander} likelihoods with a standard six-parameter $\Lambda$CDM cosmology, varying all 14 nuisance parameters\citep{Planck2013Like}.  We compare the $1$ and $2$-dimensional marginalised posteriors of the $6$ $\Lambda$CDM parameters. \CosmoChord{} is in close agreement with the posteriors produced by \CosmoMC{}, recovering the correct mean values of and degeneracies between the parameters. \label{fig:cosmochord}}
\end{figure*}

An additional strength of \PolyChord{} lies in its ability to exploit a fast-slow hierarchy common in many cosmological applications. We have successfully implemented \PolyChord{} within \CosmoMC{}, and term the result \CosmoChord{}.  The traditional Metropolis--Hastings algorithm is replaced with nested sampling. This implementation is available to download from the link at the end of the paper.

The exploitation of fast-slow parameters means that \CosmoChord{} vastly outperforms \MultiNest{} when running with modern Planck likelihoods. 

\CosmoMC{} by default uses a Metropolis--Hastings sampler. If this has a well-tuned proposal distribution (e.g.\ if one is performing importance sampling from an already well-characterised likelihood), then \PolyChord{} is $2$--$4$ times slower than the traditional \CosmoMC{}. If proposal matrices are unavailable (e.g.\ in the case that one is examining an entirely new model) then \CosmoChord{}'s run time is significantly faster than the native \CosmoMC{} sampler. This is a good example of the self-tuning capacity of \PolyChord{}, since it only requires two tuning parameters, as opposed to $\bigO{D}$.

\CosmoChord{} produces parameter estimations consistent with \CosmoMC{} (Figure~\ref{fig:cosmochord}).
It has been implemented effectively in multiple cosmological applications in the latest Planck paper describing constraints on inflation~\citep{planck2015-a24}, including application to a $37$-parameter reconstruction problem ($4$ slow, $19$ semi-slow, $14$ fast). 
In addition, \PolyChord{} is an integral component of the \ModeChord{} code, a combination of \CosmoChord{} and \ModeCode{} \citep{ModeChord1,ModeChord2,ModeChord3}, which is available at \url{http://modecode.org/}.

\section{Conclusions}
\label{sec:conclusions}
We have introduced \PolyChord{}, a novel nested sampling algorithm tailored for high-dimensional parameter spaces. It is able to fully exploit a hierarchy of parameter speeds such as is found in \CosmoMC{} and \CAMB{}~\citep{cosmomc,CAMB}. It utilises slice sampling at each iteration to sample within the hard likelihood constraint of nested sampling. It can identify and evolve separate modes of a posterior semi-independently and is parallelised using \openMPI{}.

\section*{Acknowledgements}
We would like to thank Farhan Feroz for numerous helpful discussions during the inception of the PolyChord algorithm.
W H thanks STFC for their support.

\section*{Download Link}
PolyChord is available for download from: \url{http://ccpforge.cse.rl.ac.uk/gf/project/polychord/}

\appendix
\section{Prior transformations}
\label{app:prior_tranformations}
Here we give examples of the procedure for calculating the transformation from the unit hypercube to the physical space. We demonstrate it for a simple separable case, and a more complicated dependent case

To recap, we aim to compute the inverse of the functions $F_i$: 
\begin{equation}
  F_i(\theta_i|\theta_{i-1},\ldots,\theta_0) = \int\limits_0^{\theta_i} \pi_i(\theta_i^\prime|\theta_{i-1},\ldots,\theta_1) d\theta_i^\prime,
  \label{eqn:appFi}
\end{equation}
where:
\begin{equation}
  \pi_i(\theta_i|\theta_{i-1},\ldots,\theta_0) 
  =
  \frac{%
    \int \pi_i(\params) d\theta_{i+1}\ldots d\theta_{N}
  }{%
    \int \pi_i(\params) d\theta_{i}\ldots d\theta_{N}
  }.
  \label{eqn:apppii}
\end{equation}
$\bFF$ maps from $\params$ in the physical space onto the unit hypercube injectively.

\subsection{Separable priors}
\label{app:separable_priors}
A separable prior satisfies:
\begin{equation}
  \pi(\params) = \prod_i\pi_i(\theta_i).
  \label{eqn:separability}
\end{equation}
This has the fortunate side effect that the functions $F_i$ only depend on $\theta_i$:
\begin{equation}
  F_i(\theta_i|\theta_{i-1},\ldots,\theta_0) = F_i(\theta_i).
\end{equation}

Solving a separable prior thus amounts to solving a one-dimensional inverse-transform sampling problem. We demonstrate this procedure for two cases, a rectangular uniform prior, and a Gaussian prior.

\subsubsection{Uniform prior}
\label{app:uniform_prior}
\newcommand{\thetamin}{\theta_\smin} 
\newcommand{\thetamax}{\theta_\smax} 
A rectangular uniform prior is defined by two parameters, ${\thetamin,\thetamax}$:
\begin{equation}
  \pi(\theta) = 
  \left\{
    \begin{array}{rl}
      {(\thetamax - \thetamin)}^{-1} 
      &
      \text{for }\thetamax<\theta_i<\thetamin \\
      0 & \text{otherwise.}
    \end{array}
  \right.
\label{eqn:uniform_prior}
\end{equation}

Computing $F(\theta)$ we find:
\begin{align}
  F(\theta) &= \int_{-\infty}^\theta \pi(\theta')d\theta', \nonumber\\
  &= \frac{\theta-\thetamin}{\thetamax-\thetamin},
  \label{eqn:calc_uniform_trans}
\end{align}
with $F=0$ or $1$ either side of $\thetamin$ and $\thetamax$ respectively. Inverting the equation $F(\theta)=x$ we find:
\begin{equation}
  \theta = \thetamin+(\thetamax-\thetamin)x,
  \label{eqn:uniform_trans}                           
\end{equation}
is the transformation from $x$ in the unit hypercube to $\theta$ in the physical space.

\subsubsection{Gaussian prior}
\label{app:gaussian_prior}
Defining a Gaussian prior with mean $\mu$ and standard deviation $\sigma$:
\begin{equation}
  \pi(\theta) = \frac{1}{\sqrt{2\pi}\sigma}\exp{\left[-\frac{{(x-\mu)}^2}{2\sigma^2}\right]},
  \label{eqn:gaussian_prior}
\end{equation}
We find that the procedure above yields:
\begin{equation}
  \theta = \mu + \sqrt{2}\sigma\text{erfinv}(2x-1),
  \label{eqn:gaussian_trans}                           
\end{equation}
where $\text{erfinv}$ is the conventional inverse error function.

\subsection{Forced identifiability priors}
\label{app:forced_identifiablility}

As an example of a prior that is not separable in the parameters, we consider a forced identifiability prior. Here, $n$ parameters are distributed uniformly between $\thetamin$ and $\thetamax$, but subject to the constraint that they are ordered numerically. This is a particularly useful prior in the reconstruction of functions using a spline with movable knots~\citep{vazquez_knots,knottedsky1,knottedsky2,planck2015-a24}. In this case, the  horizontal locations of the knots must be ordered.

The required prior is uniform in the hyper-triangle defined by $\thetamin<\theta_1<\cdots<\theta_n<\thetamax$, and zero everywhere else:
\begin{equation}
  \pi(\params) = 
  \left\{
    \begin{array}{rl}
      \frac{1}{n!{(\thetamax - \thetamin)}^{n}} 
      &
      \text{for }\thetamin<\theta_1<\cdots<\theta_n<\thetamax \\
      0 &\text{otherwise.}
    \end{array}
    \right.
\label{eqn:sorted_uniform_prior}
\end{equation}

To calculate equations~(\ref{eqn:appFi} \&~\ref{eqn:apppii}) we simply integrate over the constant distribution, taking care with the limits. We find:
\begin{align}
  \pi_i(\theta_i|\theta_{i-1},\ldots,\theta_0) &= \frac{(n-i+1){(\theta_i-\theta_{i-1})}^{n-i}}{{(\thetamax-\thetamin)}^{n-i+1}},\\
  F_i(\theta_i|\theta_{i-1},\ldots,\theta_0) &= {\left(\frac{\theta_i-\theta_{i-1}}{\thetamax-\theta_{i-1}}\right)}^{n-i+1},
  \label{eqn:sorted_uniform_calc}
\end{align}
where for consistency we define $\theta_0 = \thetamin$. Hence solving $x_i=F(\theta_i|\theta_{i-1},\ldots,\theta_0)$ for $\theta_i$ we find:
\begin{equation}
  \theta_i = \theta_{i-1}+ (\thetamax-\theta_{i-1})x_i^{1/(n-i+1)}.
  \label{eqn:sorted_uniform_trans}
\end{equation}
This enables $\{\theta_i\}$ to be calculated sequentially from $\{x_i\}$. We may interpret this transformation as $\theta_i$ being distributed as the smallest of $n-i+1$ uniformly distributed variables in the range $[\theta_{i-1},\thetamax]$.

\section{Evidence estimates and errors}                            
\label{app:evidences}

\cite{skilling2006} initially advocated using Monte-Carlo methods to estimate the evidence error, although this requires the storage of the entire chain of dead points, rather than just the subset usually stored for posterior inferences. For high-dimensional problems, the number of dead points is prohibitively large, and cannot be stored.

\cite{MultiNest2} use an alternative method based on the relative entropy (also suggested by~\cite{skilling2006}). 

\cite{Keeton} suggests a more intuitive methodology of estimating the error, and it is this which we use, although it must be heavily adapted for the case of variable numbers of live points and clustering.

\subsection{Basic theory}
\label{app:basic_theory}

We wish to compute the sum:
\begin{equation}
  \ev = \sum\limits_{i} (X_{i-1}-X_{i})\lik_i.
  \label{eqn:app_ev}
\end{equation}
However, we do not know the volumes $X_i$ exactly, so we can only make inferences about $\ev$, in terms of a probability distribution $\Prob(\ev)$. In practice, all we need to compute is the mean and variance of this distribution:
\begin{align}
  \text{mean}(\ev) &\equiv \overline{\ev},\\
  \text{var}(\ev) &\equiv \overline{\ev^2}-\overline{\ev}^2.
\end{align}
At iteration $i$, the $\nlive$ live points are each uniformly sampled within a contour of volume $X_{i-1}$. The volume $X_i$ will be the largest volume out of $\nlive$ uniform volume samples in volume $X_i$.
Thus $X_i$ satisfies the recursion relation:
\begin{align}
  X_i &= t X_{i-1}, \qquad X_0=1, \label{eqn:rr_x} \\
  P(t) &= \nlive t^{\nlive - 1}, \label{eqn:Pt}
\end{align}
where the $t$ and $X_{i-1}$ are independent.

It is worth noting that the procedure described below will generate the mean and variance of the distribution, but in fact this is not quite what we want. The evidence is in practice approximately log-normally distributed. Thus, it is better to report the mean and variance of $\log\ev$, defined by:
\begin{align}
  \text{mean}(\log\ev) &= 2\log\overline{\ev} - \frac{1}{2}\log\overline{\ev^2},\\
  \text{var}(\log\ev) &= \log\overline{\ev^2}-2\log\overline{\ev}.
\end{align}

\subsection{Computing the mean evidence}
\label{app:basic_mean}

While it is possible to take equations~(\ref{eqn:app_ev},\ref{eqn:rr_x} \&~\ref{eqn:Pt}) and compute the mean as a general formula~\citep{Keeton}, in the case of clustering this is uninformative. 
In fact, for large-dimensional spaces using the full formula would require storage of a prohibitively large amount of data. The calculation is better accomplished by a set of recursion relations, which update the mean evidence and its error at each step. 

For now, assume that we have $n$ live points currently enclosed by some likelihood contour $\lik$ of volume $X$, and $\ev$ is the last value of the evidence calculated from all of the points that have died so far. By considering~(\ref{eqn:app_ev},\ref{eqn:rr_x}\&\ref{eqn:Pt}), when we kill off the outermost point, we may adjust the values of $\ev$ and $X$ using:
\begin{align}                                                         
\ev &\to \ev + (1-t)X\lik,
\label{eqn:rr_Z}
\\
X &\to tX.
\label{eqn:rr_X}
\end{align}
Taking the mean of these relations, we may use the facts that $t$ and $X$ are independent random variables and that $P(t) = n t^{n-1}$, to find the recursion relations:
\begin{align}
  \overline{\ev} &\to \overline{\ev} + \frac{1}{n+1}\overline{X}\lik,
  \label{eqn:rr_Zb}
  \\
  \overline{X} &\to \frac{n}{n+1}\overline{X}.
  \label{eqn:rr_Xb}
\end{align}

\subsection{Computing the evidence error}
\label{app:basic_error}
To estimate $\overline{\ev^2}$, we square~(\ref{eqn:rr_Z}) and~(\ref{eqn:rr_X}) and multiply both together to obtain:
\begin{align}
  \ev^2 &\to \ev^2 + 2(1-t)\ev X\lik +  {(1-t)}^2X^2\lik ^2,
  \label{eqn:rr_Z2b}
  \\
  \ev X &\to t\ev X + t(1-t)X^2\lik,
  \label{eqn:rr_ZXb}
  \\
  X^2 &\to t^2X^2.
  \label{eqn:rr_X2b}
\end{align}
Note that we now need to keep track of the variable $\ev X$, as these two are not independent.
Taking the averages of the above yields:
\begin{align}
  \overline{\ev^2} &\to \overline{\ev^2} + \frac{2\overline{\ev X}\lik}{n+1} +  \frac{\overline{X^2}\lik^2}{(n+1)(n+2)},
  \\
  \overline{\ev X} &\to \frac{n\overline{\ev X}}{n+1} + \frac{n\overline{X^2}\lik}{(n+1)(n+2)},
  \\
  \overline{X^2} &\to \frac{n}{n+2}\overline{X^2}.
\end{align}

\subsection{The full calculation}
\label{app:basic_full}

There are therefore five quantities to keep track of: 
\[ \overline{\ev},\quad\overline{\ev^2},\quad\overline{\ev X},\quad \overline{X},\quad\overline{X^2}.\]
These should be initialised at $\{0,0,0,1,1\}$ respectively, and updated using equations~(\ref{eqn:rr_Zb},\ref{eqn:rr_Z2b},\ref{eqn:rr_ZXb},\ref{eqn:rr_Xb},\ref{eqn:rr_X2b}) in that order. In fact, we keep track of the logarithm of these quantities, in order to avoid machine precision errors.

\section{Evidence estimates and errors in clusters}
\label{app:evidences_clusters}
This analysis follows that of Appendix~\ref{app:evidences}. We recommend that you have understood the methods described there before continuing.

Throughout the algorithm, there will in general be $m$ identified clusters. In doing so, we wish to keep track of the volume of each cluster $\{X_1,\ldots,X_m\}$, the global evidence and its error $\ev,\ev^2$ and the local evidences and their errors $\{\ev_1,\ev_1^2\ldots,\ev_m,\ev_m^2\}$. At each iteration, the point with the lowest likelihood $\lik$ will be killed from cluster $p$, ${(1\le p\le m)}$.

\subsection{Evidence}
\label{app:cluster_ev}

We thus need to update the global evidence, the local evidence of cluster $p$, and the volume of cluster $p$:
\begin{align}
  \ev &\to \ev + (1-t)X_p\lik,
  \label{eqn:rr_ev}\\
  \ev_p &\to \ev_p + (1-t)X_p\lik,
  \label{eqn:rr_evi}\\
  X_p &\to t X_p.
  \label{eqn:rr_Xi}
\end{align}
Since $t$ will be distributed with $P(t) = n_p t^{n_p-1}$,
taking the mean of these yields:
\begin{align}
  \overline\ev &\to \overline\ev + \frac{\overline X_p \lik}{n_p+1},\\
  \overline\ev_p &\to \overline\ev_p + \frac{\overline X_p \lik}{n_p+1},\\
  \overline X_p &\to \frac{n_p\overline X_p }{n_p+1}. 
\end{align}
Keeping track of $\{\overline{\ev},\overline{\ev_p},\overline{X_p},p=1\ldots m\}$ and updating them using the recursion relations in the order above will produce a consistent evidence estimate for both the local and global evidence errors.

\subsection{Evidence errors}
\label{app:cluster_err}

We must also keep track of the local and global evidence errors. Taking the square of equations~(\ref{eqn:rr_ev} \&~\ref{eqn:rr_evi}) yields:
\begin{align}
  \ev^2 &\to \ev^2 + 2 (1-t) \ev X_p \lik +  {(1-t)}^2 X_p^2 \lik^2, \\
  \ev_p^2 &\to \ev_p^2 + 2 (1-t) \ev_p X_p \lik +  {(1-t)}^2 X_p^2 \lik^2.
\end{align}
We can see that we're going to need to keep track of $\{\overline{\ev X_p}, \overline{\ev_p X_p}, \overline{X_p^2}\}$ in addition to $\{ \overline{\ev^2}, \overline{\ev_p^2} \}$. Taking various multiplications of equations~(\ref{eqn:rr_ev},~\ref{eqn:rr_evi} \&~\ref{eqn:rr_Xi}) finds:
\begin{align}
  \ev X_p   &\to t \ev X_p + (1-t)t X_p^2 \lik, \\
  \ev X_q   &\to   \ev X_q + (1-t) X_p X_q \lik \qquad (p\ne q),   \\
  \ev_p X_p &\to t \ev_p X_p + (1-t)t X_p^2 \lik, \\
  X_p^2     &\to t^2 X_p^2, \\
  X_p X_q   &\to t   X_p X_q.
\end{align}
Taking the mean of the above yields the recursion relations:
\begin{align}
  \overline{\ev^2} &\to \overline{\ev^2} + \frac{2\overline{\ev X_p}\lik_p}{n_p+1}  + \frac{2\overline{X_p^2}\lik^2}{(n_p+1)(n_p+2)}, \\
  \overline{\ev_p^2} &\to \overline{\ev_p^2} + \frac{2\overline{\ev_p X_p}\lik}{n_p+1}  + \frac{2\overline{X_p^2}\lik^2}{(n_p+1)(n_p+2)}, \\
  \overline{\ev X_p} &\to \frac{n_p\overline{\ev X_p}}{n_p+1}  + \frac{n_p\overline{X_p^2} \lik}{(n_p+1)(n_p+2)},   \\
  \overline{\ev X_q} &\to \overline{\ev X_p}  + \frac{\overline{X_p X_q} \lik}{(n_p+1)} \qquad (q\ne p),  \\
  \overline{\ev_p X_p} &\to \frac{n_p\overline{\ev_p X_p}}{n_p+1}  + \frac{n_p\overline{X_p^2} \lik}{(n_p+1)(n_p+2)},   \\
  \overline{X_p^2} &\to \frac{n_p\overline{X_p^2}}{n_p+2}, \\
  \overline{X_p X_q} &\to \frac{n_p\overline{X_p X_q}}{n_p+1}.
\end{align}
Keeping track of 
\[\{\overline{\ev^2},\overline{\ev_p^2},\overline{\ev X_p},\overline{\ev_p X_p},\overline{X_p^2},\overline{X_p X_q},p,q=1\ldots m\},\]
and updating them using the recursion relations in the order above will produce a consistent estimate for the local and global evidence errors.

\subsection{Cluster initialisation}
\label{app:cluster_init}
All that remains is to initialise the clusters correctly at the point of creation.

The starting initialisation of the evidence and volume is reasonable, there will be only a single cluster with volume $1$, and all evidence related terms $0$. At some point (possibly at the beginning, depending on the prior), the live points will split into distinct clusters, and the local volumes and evidences will need to be re-initialised.

At the point of splitting a cluster into sub-clusters, we partition the $n$ live points into a $N$ new clusters, with $\{n_1,\ldots,n_N\}$ live points in each. If the volume of the splitting cluster is $X_p$ initially, we need to know how to partition this volume into $\{X_1,\ldots,X_N\}$. If the points are drawn uniformly from the volume, then the $n_i$ will depend on the volumes via a multinomial probability distribution:
\begin{equation}
  P(\{n_i\}|X_p,\{X_i\}) \propto {X_1}^{n_1} \ldots X_N^{n_N}.
\end{equation}
We however want to know the probability distributions of the $\{X_i\}$, given the $\{n_i\}$. We can invert the above with Bayes' theorem, using an (improper) logarithmic prior on the volumes subject to the constraint that they sum to $X_p$:
\begin{equation}
  P(\{X_i\}|X_p) \propto \frac{\delta(X_1+\cdots+X_N-X_p)}{X_1\cdots X_N}.
\end{equation}
Doing this shows the posterior $P(\{X_i\}|X_p,\{n_i\})$ is a Dirichlet distribution with parameters $\{n_i\}$. More importantly, we can use this to compute the means and correlations for the volumes $\{X_i\}$:
\begin{align}
  \overline{X_i}&=  \frac{n_i}{n} \overline X_p, \\
  \overline{X_i^2}&= \frac{n_i(n_i+1)}{n(n+1)} \overline{X_p^2}, \\
  \overline{X_i X_j} &= \frac{n_i n_j}{n(n+1)} \overline{X_p^2}, \\
  \overline{X_i Y} &= \frac{n_i}{n} \overline{X_p Y} \qquad Y\in \{Z,Z_p,X_q\}.
\end{align}
The first equation recovers the intuitive result that the volume should split as the fraction of live points. Note, however that this requires a logarithmic prior. The third shows us that since $\overline{X_i X_j}\ne\overline{X_i}\:\overline{X_j}$, the volumes are correlated at the splitting. This is to be expected.

We also need to initialise the local evidences and their errors. A consistent approach is to assume that the evidences also split in proportion to the cluster distribution of live points. Following the same reasoning as above, we find that:
\begin{align}
  \overline{Z_i}&=  \frac{n_i}{n} \overline Z_p \\
  \overline{Z_i X_i}&= \frac{n_i(n_i+1)}{n(n+1)} \overline{Z_p X_p} \\
  \overline{Z_i^2}  &= \frac{n_i(n_i+1)}{n(n+1)} \overline{Z_p^2} \\
\end{align}

Thus, at cluster splitting, all of the new local evidences, volumes and cross correlations are initialised according to the above.

This completes the mechanism for keeping track of the local and global evidences, their errors, and the local cluster volumes.

\label{lastpage} 

\bibliographystyle{apj}
\bibliography{polychord}

\end{document}